\documentclass{aa}  

\usepackage{array,multirow,graphicx}
\usepackage{natbib}
\usepackage{gensymb}
\usepackage{dblfloatfix}
\usepackage{multicol}
\usepackage{blindtext}
\usepackage{float}
\usepackage{subfig}
\usepackage{hyperref}
\usepackage{geometry,graphicx,kantlipsum,rotating}
\usepackage{ulem}
\bibpunct{(}{)}{;}{a}{}{,} 

\hypersetup{
    colorlinks=true,                
    breaklinks=true,                
    urlcolor= red,                  
    linkcolor= blue,                
    citecolor= blue                 
    }
\usepackage{graphicx}
\usepackage{txfonts}
\usepackage[switch]{lineno}

\RequirePackage{soul}
\RequirePackage{xcolor}
\RequirePackage{todonotes}

\begin{document}

   \title{Flux variability from ejecta in structured relativistic jets with large-scale magnetic fields}

\titlerunning{Flux variability from ejecta in structured relativistic jets.}

   \author{G. Fichet de Clairfontaine
          \inst{1}
          Z. Meliani
          \inst{1}
          A. Zech
          \inst{1}
          \and 
          O. Hervet
          \inst{2}
          }

\authorrunning{G. Fichet de Clairfontaine, Z. Meliani, A. Zech, O.Hervet}

   \institute{LUTH, Observatoire de Paris, CNRS, PSL, Universit\'{e} de Paris; 5 Place Jules Janssen, 92190 Meudon, France\\
              \email{gaetan.fichetdeclairfontaine@obspm.fr} 
         \and
            Santa Cruz Institute for Particle Physics and Department of Physics, University of California at Santa Cruz, Santa Cruz,CA 95064, USA\\
            \email{ohervet@ucsc.edu}
             }

  \abstract
  {Standing and moving shocks in relativistic astrophysical jets are very promising sites for particle acceleration to large Lorentz factors and for the emission from the radio up to the $\gamma$-ray band. They are thought to be responsible for at least part of the observed variability in radio-loud active galactic nuclei.
  }
 {We aim to simulate the interactions of moving shock waves with standing recollimation shocks in structured and magnetized relativistic jets and to characterize the profiles of connected flares in the radio light curve.}
  {Using the relativistic magneto-hydrodynamic (MHD) code \texttt{MPI-AMRVAC} and a radiative transfer code in post-processing, we explore the influence of the magnetic-field configuration and transverse stratification of an over-pressured jet on its morphology, on the moving shock dynamics, and on the emitted radio light curve. First, we investigate different large-scale magnetic fields with their effects on the standing shocks and on the stratified jet morphology. Secondly, we study the interaction of a moving shock wave with the standing shocks. We calculated the synthetic synchrotron maps and radio light curves and analyze the variability at two frequencies 1 and 15.3 GHz and for several observation angles. 
  Finally, we compare the characteristics of our simulated light curves with radio flares observed from the blazar 3C\,273 with the Owens Valley Radio Observatory (OVRO) and Very Long Baseline Array (VLBA) in the MOJAVE survey between 2008 and 2019.
   }
  {We find that in a structured over-pressured relativistic jet, the presence of the large-scale magnetic field structure changes the properties of the standing shock waves and leads to an opening in the jet.
  The interaction between waves from inner and outer jet components can produce strong standing shocks. When crossing such standing shocks, moving shock waves accompanying overdensities injected in the base of the jet cause very luminous radio flares. The observation of the temporal structure of these flares under different viewing angles probes the jet at different optical depths. 
  At 1 GHz and for small angles, the self-absorption caused by the moving shock wave becomes more important and leads to a drop in the observed flux after it interacts with the brightest standing knot. A weak asymmetry is seen in the shape of the simulated flares, resulting from the remnant emission of the shocked standing shocks. 
  The characteristics of the simulated flares and the correlation of peaks in the light curve with the crossing of moving and standing shocks favor this scenario as an explanation of the observed radio flares of 3C\,273.
  }
  {}


   \maketitle

\section{Introduction}

Emission from relativistic jets in radio-loud active galactic nuclei (AGN) has been detected from the radio band up to the teraelectronvolt range and shows frequent episodes of high flux variability in many sources. The emission is generally ascribed to a population of relativistic particles inside the jet that interact with magnetic and photon fields to produce nonthermal radiation over a large wavelength range. In blazar-type AGN, the emission from the jet plasma is amplified in the observer frame by relativistic beaming effects as the jet axis is aligned with the line of sight, while the observed emission is weaker in radio galaxies, where the jet is seen under a larger angle.
The multiwavelength emission of AGN requires a mechanism able to reaccelerate particles as they travel along the jet \citep{Blandford_1979ApL....20...15B}. The processes that are most frequently proposed are acceleration from internal shocks \citep{Lemoine_2019, Lemoine_2019_2, Lemoine_2019_3}, shear acceleration \citep{Rieger_2019, Tavecchio_2020}, or magnetic reconnection \citep{Blandford_2017}. 

Many direct and indirect observations demonstrate the presence of a transverse stratified profile of AGN jets, characterized by the presence of a fast inner jet (spine) and a slower outer jet (sheath or layer), both inner and outer jets being relativistic. 
The most compelling observation of this structure at the limb-brightened jets was observed with very-long-baseline interferometry (VLBI), down to a few Schwarzschild radii for the nearest radio galaxies \citep{Nagai_2014, Kim_2018}. Seen at large angles, the slower layer is more Doppler boosted than the spine, leading to the appearance of a distinctive radio structure of an "empty pipe."
Observations of a different magnetic structure of the inner and outer jets via polarization measurements \citep{Gabuzda_2014, Avachat_2016, Giroletti_2004} support the idea that the two jet components may be launched from different processes, such as those proposed by \cite{Blandford_1977} and \cite{Blandford_1982}, for launching from the vicinity of the rotating supermassive black hole or from the accretion disk.

Theoretical studies in plasma physics also support this interpretation where the fast inner jet is responsible for most of the radiative output while having a lower density and a population dominated by electrons and positrons, while the outer jet is denser and less radiative, dominated by cold protons \citep{Sol_1989}.
The gamma-ray detection of radio galaxies is challenging to explain when considering a uniform jet. It might be better explained by a stratified jet structure, where the particle and synchrotron fields of both jet components interact to produce a strong high-energy inverse-Compton emission \citep{Tavecchio_2008, Tavecchio_2014}.
The stratification of the relativistic jet can explain the spectral shape of the emission at multiwavelength, from the radio band to the X-ray band \citep{Siemiginowska_2007}, and up to the (very) high energy gamma-ray band \citep{Ghisellini_2005}.

Large-scale magnetic fields seem to play an important role in extragalactic jets, in particular for the collimation of the jet and its acceleration \citep{Porth_2011, Fendt_2012}. Observations tend to show a correlation between the large-scale magnetic structure and the resulting synchrotron emission \citep{Gabuzda_2014, Walker_2018}.

The jet emission often shows a presence of bright spots ("knots") that can be associated with standing and moving shocks. Such features have been detected in relativistic jets with radio and optical polarimetry observations \citep{Perlman_1999, Marshall_2002}, in the radio and millimeter band \citep{Britzen_2010, Jorsatd_2013} and in the X-ray band \citep{Wilson_2002}. One way to interpret these "knots" is by evoking recollimation shocks along the jet 
caused by pressure mismatches with the medium surrounding the jet \citep{Marscher_2008Natur.452..966M}.

Flux variability is a characteristic feature of emission from radio-loud AGN and it depends on the observed frequency range and the AGN class.
In the gigahertz radio bands, the shortest observed flare time scales can be of a few months in the case of BL Lac objects observed at a frequency of $\nu =43$ GHz \citep{Werhle_2012, Wehrle_2016}, to a few years in the sample of several tens of radio-loud AGN observed at frequencies from $\nu = \left[ 4.8;~230 \right]~\text{GHz}$ \citep{Hovatta_2008,Nieppola_2009}. In the latter, the median flare duration was estimated to be 2.5 years, with flares occurring on average every four to six years.
While in many cases the light curves of the detected flares exhibited a complex structure, sometimes including multiple peaks, in general, the decay times were found to be typically 1.3 times longer than the rise times. Their analysis concludes that the observed flare characteristics are in agreement with a generalized shock model \citep{Valtaoja_1992}.
In the case of the very nearby radio-galaxy M\,87, VLBI observations \citep{Walker_2018} can locate fast flux variability from the radio core. 
At high energies, in X-rays and gamma-rays, very rapid flares are observed from blazars and radio-galaxies with durations of days, down to time scales below an hour at teraelectronvolt energies. 

 Since the first analytical model \citep{Blandford_1979} that was able to reproduce the flat radio spectra of jets with an isothermal jet associated with shock waves traveling in the jet, models have been evolving in several directions. Hydrodynamic and MHD models are developed for in-depth studies of jet launching and propagation, while models of radiative transfer focus on the description of multiwavelength emission processes due to relativistic particle populations in the emission region, and particle in cell (PIC) models treat the microphysics of particle acceleration at small scales.

Today, increasingly sophisticated simulations address at the same time the macrophysics of the jet plasma and its radiative emission to try to improve our understanding of the jet physics from multiwavelength observations. Several hydrodynamical simulations of jets \citep{Gomez_1997, Agudo_2001, Mimica_2009} have shown that injection of perturbations at the base of the jet succeeds in reproducing the observed radio structure of superluminal and stationary components accounting for synchrotron radiation from a randomly oriented magnetic field. These simulations have also shown that perturbations traveling along an over-pressured jet can lead to the appearance of recollimation shocks. 

Including a large-scale magnetic field structure in simulations of relativistic jets, \cite{Mizuno_2015} studied the impact of the geometry of the magnetic field on recollimation shocks and rarefaction structures. They showed that the influence of the magnetic structure is not negligible and that, for example, axial fields lead to stronger collimation and rarefaction than toroidal fields. In studies by \cite{Marti_2015} and \cite{Marti_2018}, the authors simulated models of relativistic magnetized, axisymmetric jets with azimuthal velocity (i.e., rotation). For certain configurations, this azimuthal velocity leads to change the stationary shock-wave structure.
Thus, they obtain a standing-shock structure and compute synthetic radio maps compatible with observations of parsec-scale extragalactic jets. \cite{Fuentes_2018} are able to obtain polarization maps by computing the optically thin and linearly polarized synchrotron emission. They find that the electric vector polarization angles tend to be perpendicular to the axis of the jet between the recollimation shocks. This characteristic polarization can be compared with that obtained in VLBI observations of blazar jets. 

\cite{Porth_2015} show that both unmagnetized and magnetized jets have great stability due to interactions with the ambient medium. The difference in pressure between the jet and the ambient medium allows the jet plasma to keep a causal connection. They propose an explanation for the Fanaroff-Riley dichotomy with different pressure values leading to the appearance of different structures of recollimation shocks. 
Simulations of stratified relativistic jets \citep{Hervet_2017} show that a two-component model of jets in interaction with an interstellar ambient medium can reproduce the observed knots through the generation of standing and moving shocks inside the jets. 

Shock waves passing through a conical relativistic jet were first evoked by \cite{Marscher_1985} to interpret a flare of the quasar 3C\,273 observed in 1983 from the millimeter to the infrared band. In this scenario, superluminal radio knots naturally arise as shocked regions in the jet.
With today's increased computing capacity, \cite{Fromm_2016,Fromm_2018} have been able to reproduce typical flares observed at different wavelengths by simulating the interaction between ejecta and a recollimation shock structure. 

Several characteristics of pc-scale relativistic jets are well reproduced with current models. Nevertheless, a better comprehension of the link between the supposed recollimation shock structure and the magnetic configuration is necessary to understand the multiwavelength observations, particularly during flares. 
We aim to understand the impact of the magnetic configuration in the jet on the dynamics of a perturbation ("ejecta") at the base of the jet, which we suppose to be the cause for the observed flares. We have studied in detail its interaction with a two-component jet for different large-scale magnetic configurations, as well as the synchrotron emission during its propagation. 
The overall aim is to reproduce typical radio flare observation by the injection of such perturbations at the base of the jet and to put constraints on the magnetic field configuration and jet structure.  

We carried out special relativistic magnetohydrodynamic (SRMHD) simulations of pc-scale jets with the \texttt{MPI-AMRVAC} code \citep{Keppens_2012}, using a two-component model of relativistic jets with different magnetic configurations. Following \cite{Hervet_2017} we considered an external jet component that carries more power than the internal one, while staying within the same order of magnitude. This kind of configuration leads to the formation of a typical standing-shock structure. We consider four different magnetic configurations (defined in cylindrical coordinates) : hydrodynamic \textbf{(H)} with a turbulent magnetic field linked to the thermal pressure, toroidal \textbf{(T)} (with magnetic lines along $\varphi$), poloidal \textbf{(P)} (with magnetic lines along $z$, the jet propagation axis) and helical \textbf{(HL)} (with a magnetic pinch angle of $45 \degree$).
Synchrotron radiation is computed in post-processing, assuming the injection of relativistic electrons following \cite{Gomez_1995} and accounting for relativistic effects. 
In this way, radio light curves can be computed for the passage of ejecta in the stationary shock structure. \\

The organization of this paper is as follows. In Section~\ref{sec: SRMHD}, we briefly present the \texttt{MPI-AMRVAC} code and the numerical method used to solve the SRMHD equations, followed by a description of the numerical implementation of the two-component jet in Section~\ref{sec : the two-component model}. The structure of standing shocks that arises in the steady-state solution of this model is discussed in Section~\ref{subsec: SRMHD results} for the four different magnetic-field configurations. The introduction of ejecta leads to the perturbations of the steady-state structure is developed in Section~\ref{sec : ejecta}.
The treatment of radiative transfer and generation of synchrotron maps and light curves in post-processing is explained in Section~\ref{sec:radiative} and results are presented.
To illustrate the relevance of our scenario to explain radio flares, recent results from
observations of the blazar 3C\,273 with the VLBA and OVRO are discussed in Section~\ref{sec : 3C273} and are qualitatively compared to our simulations. Section~\ref{sec : discussion} provides a general discussion of the implications of our scenario.

Throughout this paper we use natural units where the speed of light $c=1$. As the distance unit will be the jet radius $R_{\rm jet}$, the time unit will be $R_{\rm jet}$ in the co-mobile frame or $R_{\rm jet}/\delta$ in the absolute frame (where $\delta$ is the Doppler factor).

\section{Governing SRMHD equations and numerical method}
\label{sec: SRMHD}

We perform the numerical simulation of the relativistic magnetized two-component jet model using the 2D ideal special-relativistic magnetohydrodynamics version of the finite volume code \texttt{MPI-AMRVAC} in conservation form, using high-resolution shock-capturing methods \citep{Meliani_2007,Keppens_2012}. 
It solves the governing conservation equation as in \citep{Marti&Muller_2015} with $\vec{U}$ the state vector and $\vec{F}^{\rm i}$ the associated flux vectors :

\begin{equation}
    \partial_{\rm t}\, \vec{U} + \partial_{\rm x^i}\, \vec{F}^i(\vec{U}) = 0\,,
\end{equation}
with :

\begin{equation}
    \vec{U} = \left(D,\,S^{\rm j},\,\tau,\,B^{\rm k} \right)^{\rm T}\,,
\end{equation}

\begin{equation}
\begin{aligned}
    \vec{F}^{\rm i} {} & = \Big( D v^{\rm i},\, S^{\rm j} v^{\rm i} + p \delta^{\rm ij} - b^{\rm j} B^{\rm i} / \gamma, \, \tau v^{\rm i} + p v^{\rm i} - b^{\rm 0} B^{\rm i} / \gamma,\, v^{\rm i} B^{\rm k} \\
    & - v^{\rm k} B^{\rm i} \Big)^{\rm T}\,,
\end{aligned}
\end{equation}

\noindent
where the rest-mass density is $D$, the momentum density in the j-direction $S^{\rm j}$ and the total energy density $\tau$, calculated in the absolute frame. They are given by :

\begin{align}
    & D = \rho \gamma\,, \\ 
    & S^{\rm j} = \rho h^* \gamma^2 v^{\rm j} - b^{\rm 0} b^{\rm j}\,, \\
    & \tau = \rho h^* \gamma^2 - p_{\rm tot} - \left(b^{\rm 0} \right)^2\,,
\end{align}
where $h^*\,=\, \left(1+\epsilon_{\rm the}+p_{\rm th}/\rho + b^2/\rho\right)$ with $\epsilon_{\rm the}$ the internal energy.
We note $b^{\rm i}$ the three vector magnetic field described in the co-moving frame ($\vec{B}$ and $\vec{v}$ describe the magnetic field and the velocity vector in the absolute frame) as :

\begin{align}
    & b^{\rm 0} = \gamma \vec{B}.\vec{v} \,,\\
    & b^{\rm i} = B^{\rm i} / \gamma + b^{\rm 0} v^{\rm i}\,.
 \end{align}
Finally, the Lorentz factor is given by $\gamma = 1/\sqrt{1 - v^{\rm i}v_{\rm i}}$ (with i running over the spatial indices $\left[1,2,3\right]$). \\

As a closure equation for the hydrodynamic part, we exploit the Synge equation of state \citep{Mathews_1971ApJ...165..147M,meliani_2004A&A...425..773M},

 \begin{equation}\label{Eq:Sec1_EOS}
     p = \frac{\left(\Gamma-1\right)\rho}{2}\left(\frac{\epsilon}{\rho}-\frac{\rho}{\epsilon}\right),
 \end{equation}
and the corresponding  effective polytropic index is given as \citep{Meliani_2008},
 
 \begin{equation}
     \Gamma_{\rm eff} = \Gamma - \dfrac{\Gamma - 1}{2} \left( 1 - \dfrac{1}{\epsilon} \right)\,,
 \end{equation}
with $\epsilon=\left(1+\epsilon_{\rm the}\right)$ is the specific internal energy. We fix $\Gamma = 5/3$; the effective index can vary in time and location between its relativistic ($\Gamma = 4/3$, when the thermal energy becomes larger than the mass energy) and its classical value ($\Gamma = 5/3$, when the thermal energy becomes negligible compared to the mass energy). \\

 The divergence free condition for the magnetic field is satisfied by using the divergence cleaning method described by \cite{Dedner_2002JCoPh.175..645D}.
 We use the Harten–Lax–van Leer–Contact (HLLC) Riemann solver \citep{mignone_2006} with third order reconstruction method \texttt{cada3} \citep{cada_2009}. The combination of \texttt{cada3} reconstruction (third order accurate) and HLLC flux computations is extremely robust and handles both sharp discontinuities and shock development accurately. \\
In the study of recollimation shocks, it is important to detect the shocks and distinguish them from compression waves in the numerical simulation. These internal shocks are in some cases very weak, making their detection difficult. For this purpose, in the hydrodynamics case, we use the shock detection algorithm by \cite{Zanotti_2010A&A...523A...8Z}. For the magnetized cases, we use a jump condition on the fast-magnetosonic number. We should note that in the simulation of the magnetized jet with only a toroidal-field component, the two methods converge at the vicinity of the jet axis.

\section{Two-component model of a magnetized relativistic jet  }
\label{sec : the two-component model}

In order to investigate the effect of the magnetic field configuration on the shock structure in a transverse structured jet, we use the two-component jet model proposed by \cite{Meliani_2007, Meliani_2009, Hervet_2017}. We adopt typical characteristics of a radio loud relativistic AGN jet, with a total kinetic luminosity of $L_{\rm kin} = 10^{46}~\text{erg.s}^{-1}$ \citep{Ghisellini_2014}, and the jet radius taken to be $R_{\rm jet}\,=\,0.1\,{\rm pc}$ at the parsec scale as observed in the jet of M\,87 \citep{Biretta_2002NewAR..46..239B}. 
Concerning the internal jet structure, we assume an external jet radius $R_{\rm out}\,=\,R_{\rm jet}$ and an internal jet radius $R_{\rm in} = R_{\rm out}/3$. We assume that the outer jet component carries an important fraction $f_{\rm out}=0.75$ of the total kinetic luminosity, 

\begin{equation}\label{Eq:SubS_Model_Lkin}
    L_{\rm out,\,kin}=f_{\rm out}\,L_{\rm kin}=\left(\gamma_{\rm out} h_{\rm out} - 1\right)\rho_{\rm out}\gamma_{\rm out}\pi \left(R^2_{\rm out} - R^2_{\rm in}\right) v_{\rm z, out}\,,
\end{equation}
with the remaining $L_{\rm in,\, kin} = \left(1-f_{\rm out}\right)\,L_{\rm kin}=0.25\,L_{\rm kin}$ carried by the inner jet component. \\

For the simulations performed in this paper as initial condition, we set a non-rotating, superfast, magnetosonic cylindrical jet surrounded by a uniform, unmagnetized ambient medium with high rest-mass density. The rest-mass density and the total pressure ratio of the outer jet $\left(\rho_{\rm  out},p_{\rm  out}\right)$ to the ambient medium $\left(\rho_{\rm am} = 10^3~\text{cm}^{-3},p_{\rm am} = 1~\text{dyn.cm}^{-2}\right)$ is $\left(\eta_{\rm \rho}=\rho_{\rm  out}/\rho_{\rm am}\,=\, 10^{-2},\eta_{\rm out, p}=p_{\rm jet,\, out}/p_{\rm am}\,=\, 1.0\right)$. We assume a more  over-pressured inner jet, with a lower rest-mass density and total pressure ratio of the inner jet $\left(\rho_{\rm  in},p_{\rm  in}\right)$ to the ambient medium $\left(\rho_{\rm am},p_{\rm am}\right)$ of $\left(\eta_{\rm \rho}=\rho_{\rm  in}/\rho_{\rm am}\,=\, 10^{-3},\eta_{\rm in, p}=p_{\rm  in}/p_{\rm am}\,=\, 1.5\right)$ \citep{Gomez_1995,Gomez_1997}. 
Moreover, the inner jet is assumed to be faster ($\gamma_{\rm  in} = 10$) than the outer jet ($\gamma_{\rm  out} = 3$) \citep{Giroletti_2004}. \\

To investigate the effects of the magnetic field on the recollimation shocks and therefore on the evolution of the ejecta, we have considered different topologies: hydrodynamic (\textbf{H}) (as reference case), toroidal (\textbf{T}), axial (poloidal) (\textbf{P}) and helical (\textbf{HL}). The jet magnetization is set through the local maximum magnetization parameters at the inner and the outer jet component. The magnetization parameter is given by,  

\begin{equation}\label{SubS_Model_Sigma}
\sigma_{\rm M}=\frac{\vec{B}^2+\left(\vec{v}\cdot\vec{B}\right)^2/2}{\rho\,h}\,.
\end{equation}
In all magnetized cases, the magnetization parameter is set $\sigma_{\rm M,\, in} = 10^{-3}$ for the inner jet component and $\sigma_{\rm M,\, out} = 10^{-4}$ for the outer jet component, sufficiently low to allow the Fermi 1 acceleration mechanism to be efficient \citep{Lemoine_2010, Sironi_2013,Plotnikov_2018}. In all cases the relativistic jet is kinematically dominated.\\

In the poloidal field case (\textbf{P}), the magnetic field is uniform and parallel to the jet flow in each component, and $\left(\sigma_{\rm M,\, in}, \sigma_{\rm M,\, out}\right)$ are constants. In the toroidal and helical cases (\textbf{T}, \textbf{HL}), we adopt the same profile as in \cite{Meliani_2009},

\begin{equation}\label{Eq:SubS_Model_Bphi}
  B_{\varphi} =  \left \{
  \begin{aligned}
  &B_{\rm \varphi, in, 0}\,\left(\frac{R}{R_{\rm \, in}}\right)^{a_{\rm in}/2} &&; \text{if}\ R < R_{\rm \, in}\,, \\
  &B_{\rm \varphi, out, 0}\,\left(\frac{R}{R_{\rm \, in}}\right)^{a_{\rm out}/2} &&; \text{if}\ R \ge R_{\rm \, in} \,,
  \end{aligned} \right.
\end{equation}
$\left(B_{\rm \varphi, in, 0},\,B_{\rm \varphi, out, 0}\right)$ are respectively the toroidal magnetic field strength of the inner and the outer jet component, at their interface and they are deduced from  (Eq.~\ref{SubS_Model_Sigma}) and we fix the exponents $(a_{\rm in},\,a_{\rm out}) = (0.5, -2)$ as in \citep{Meliani_2009}. It should be noted that the value of these exponents can have an influence on the resulting pattern of the recollimation shocks and the moving shock.

Concerning the helical-field case (\textbf{HL}), we chose the same configuration as for the toroidal-field case (Eq.~\ref{Eq:SubS_Model_Bphi}) and a constant axial field strength with constant magnetic pitch angle $\theta_{B} = 45^{\degree}$. \\

Finally, the thermal pressure profile $p_{\rm th}(r)$ is deduced by assuming for each component a transverse equilibrium among the pressure gradient and Lorentz force following \citep{Meliani_2009},

\begin{equation}
    p_{\rm th}(r) = p_{\rm tot} -   \left( \left(1 - v_{\rm z}^2 \right)\cdot \left(\dfrac{1}{a_{\rm in}+0.5}\right) \cdot\dfrac{B_\varphi^2}{2 } + \dfrac{B_{\rm z}^2}{2}\right)\,.
    \label{eq: thermal pressure}
\end{equation}

The initial distributions of $B_\varphi$ and $p_{\rm mag}$ in the different jet components can be seen in Fig. \ref{fig: Additional figs}.

The smooth transition between the two components of the jet is imposed using a hyperbolic tangential function with an extension of $R_{\rm  \, in}/2$ on the density, and magnetization parameter $\sigma_{\rm M}$.
We note that all physical parameters are calculated with respect to the relativistic jet kinetic energy flux, radius, Lorentz factor and magnetization. 

To make the simulation with high resolution and large spatial extend more tractable, we assume axisymmetry. The simulations are carried out in a domain size $\left[R, Z\right]= \left[16, 200\right] ~R_{\rm jet}$; we take a base resolution of $64\times 1000$ and we allow for five levels of refinement, achieving an effective resolution of $1024\times 16000$. For the initial state, the jet inlet is set over the full jet axis $Z$.

\section{Results from steady-state jet simulations}
\label{subsec: SRMHD results}

In the following, we present simulation results for the four magnetic configurations (\textbf{H}, \textbf{T}, \textbf{P}, \textbf{HL}). The simulations are run until they reach a steady state with a stationary shock pattern appearing within the two components of the jet over the full simulation box (Fig. \ref{fig: 2D contour energy, Lorentz factor - H T P HL}).

\begin{sidewaysfigure*}
\centering

\begin{multicols}{2}

    \includegraphics[width = \columnwidth]{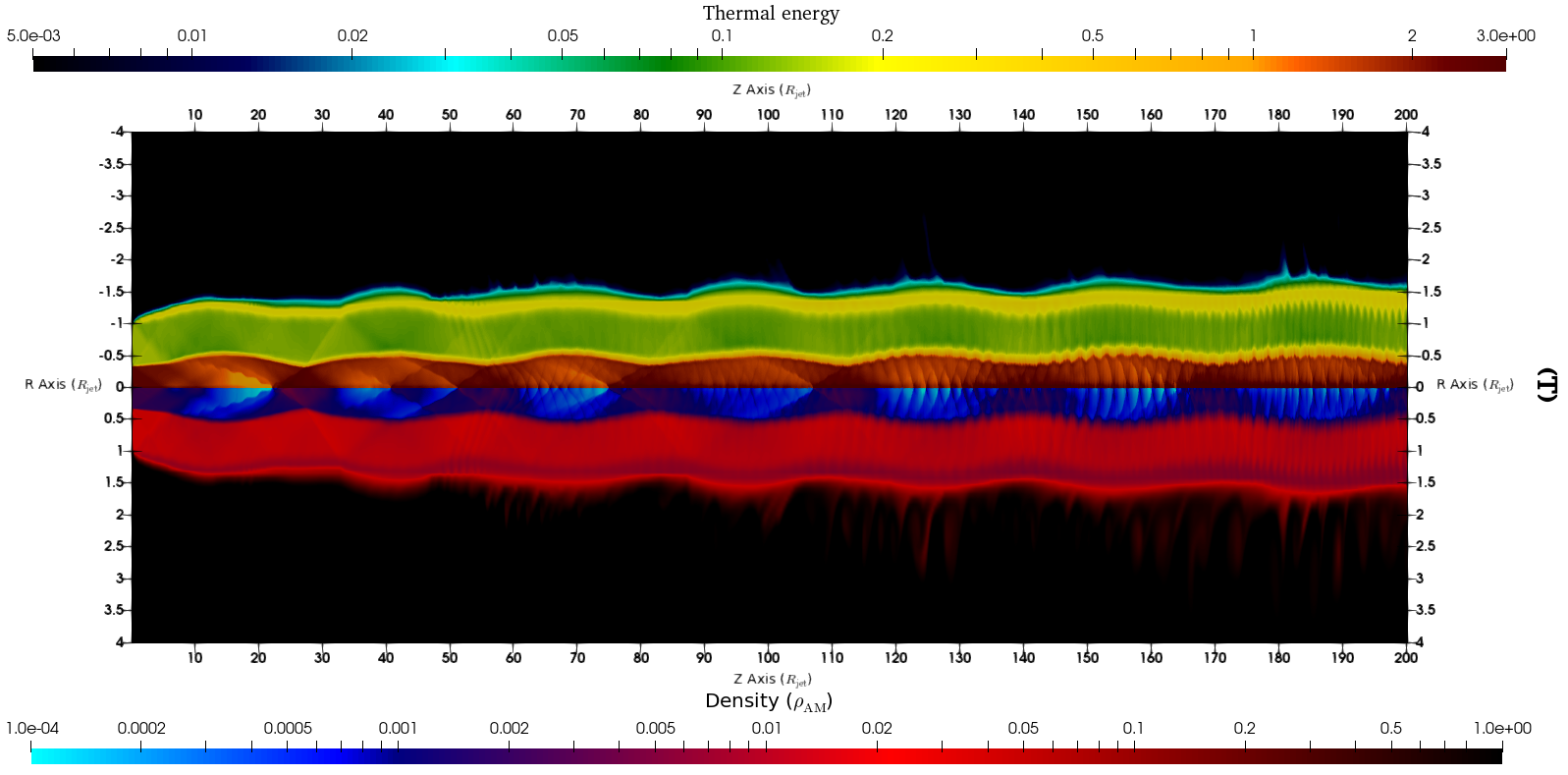}

    \includegraphics[width = \columnwidth]{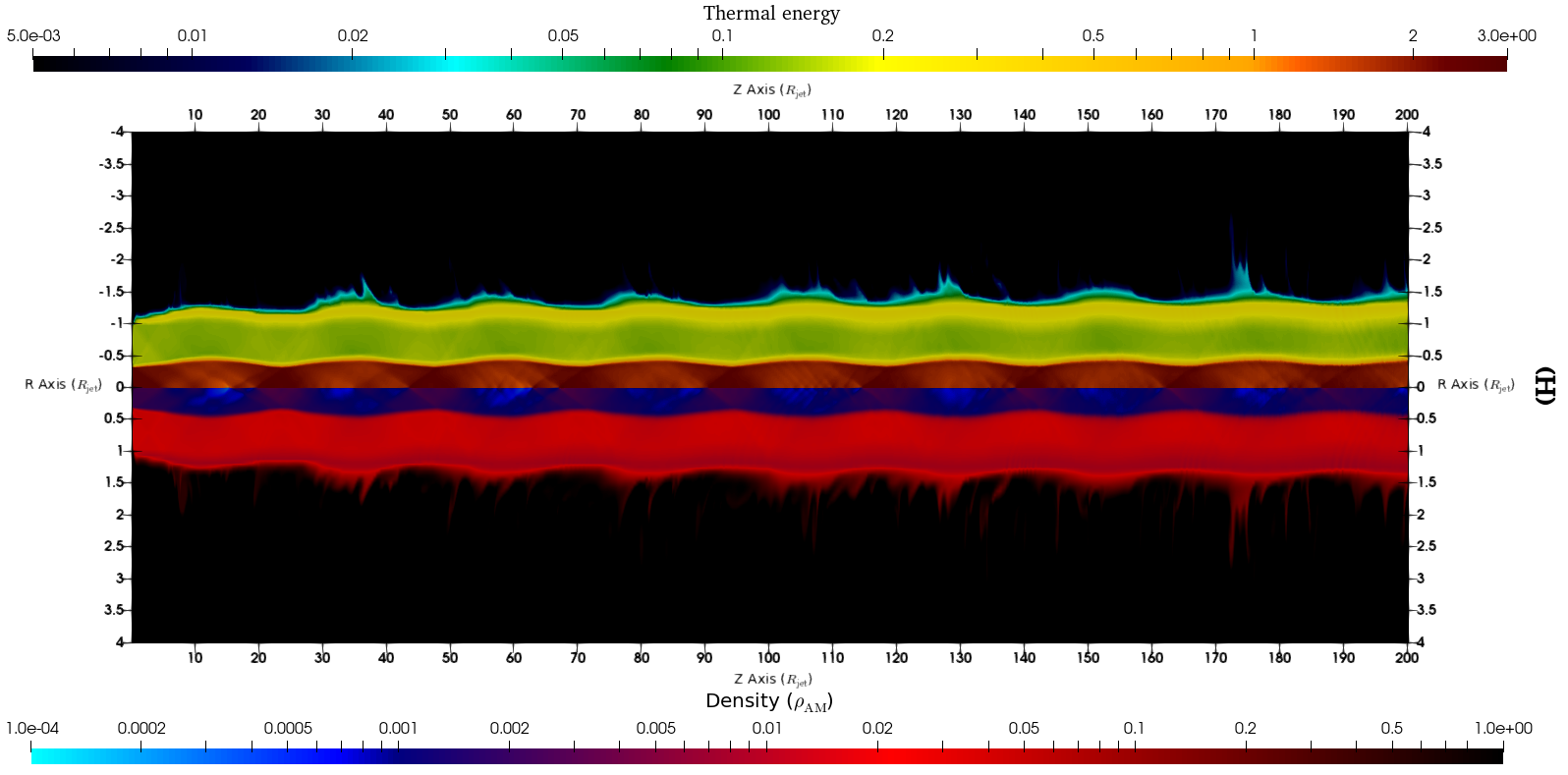}
    
\end{multicols}
\begin{multicols}{2}

    \includegraphics[width = \columnwidth]{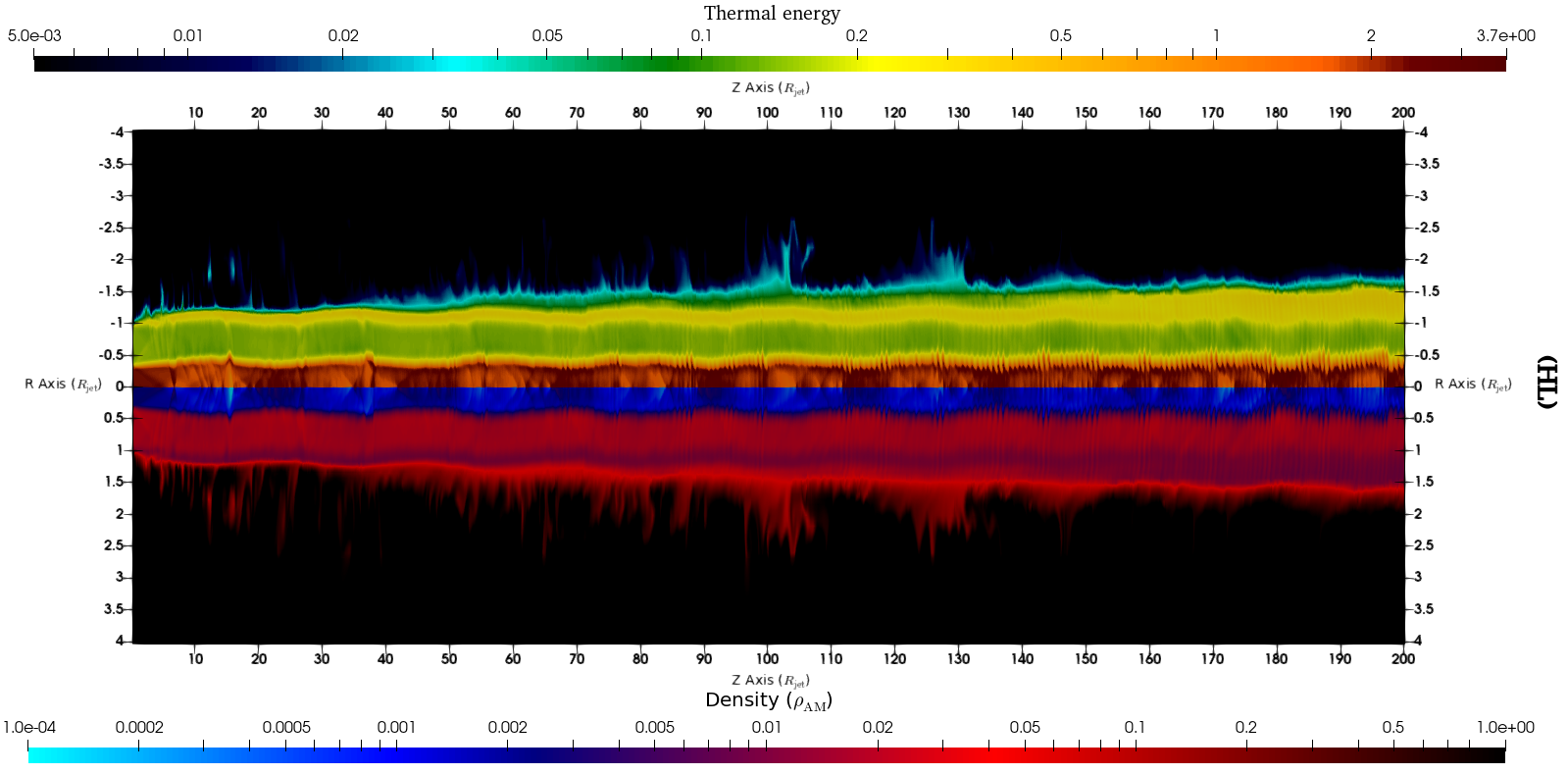}

    \includegraphics[width = \columnwidth]{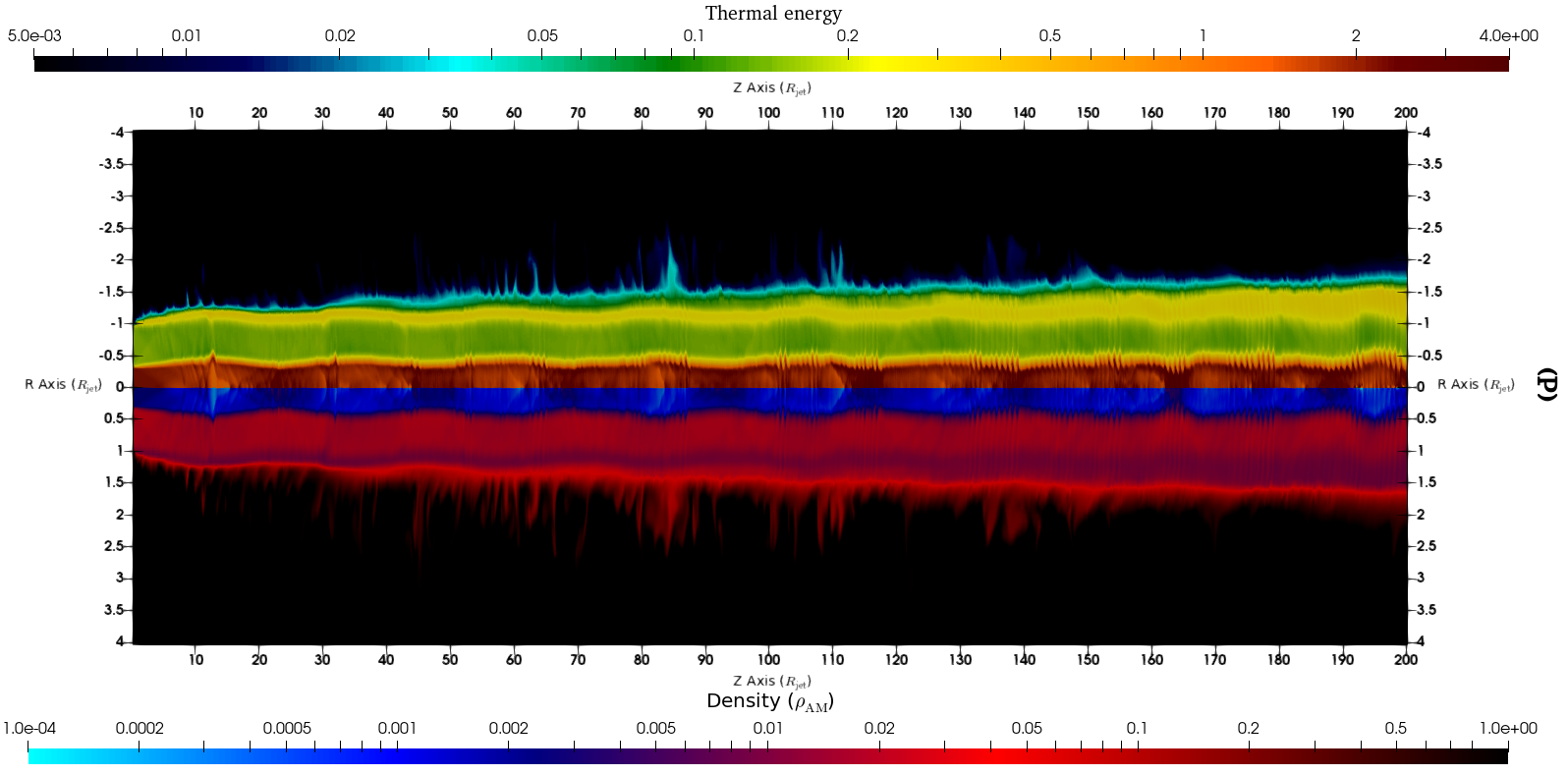}

\end{multicols}

    \caption{Snapshot: different types of structured jets (\textbf{H}, \textbf{T}, \textbf{P} and \textbf{HL}) without an injected perturbation. The propagation of the jet is going from left to right along the increasing value of Z. For each case, the density contour (in log-scale) is drawn on the bottom side and the thermal energy contour on the top side. Units on x and y-axis in $R_{\rm jet}$ unit.}
    \label{fig: 2D contour energy, Lorentz factor - H T P HL}

\end{sidewaysfigure*}

\begin{figure*}[h]
    \centering
    \includegraphics[width=2\columnwidth]{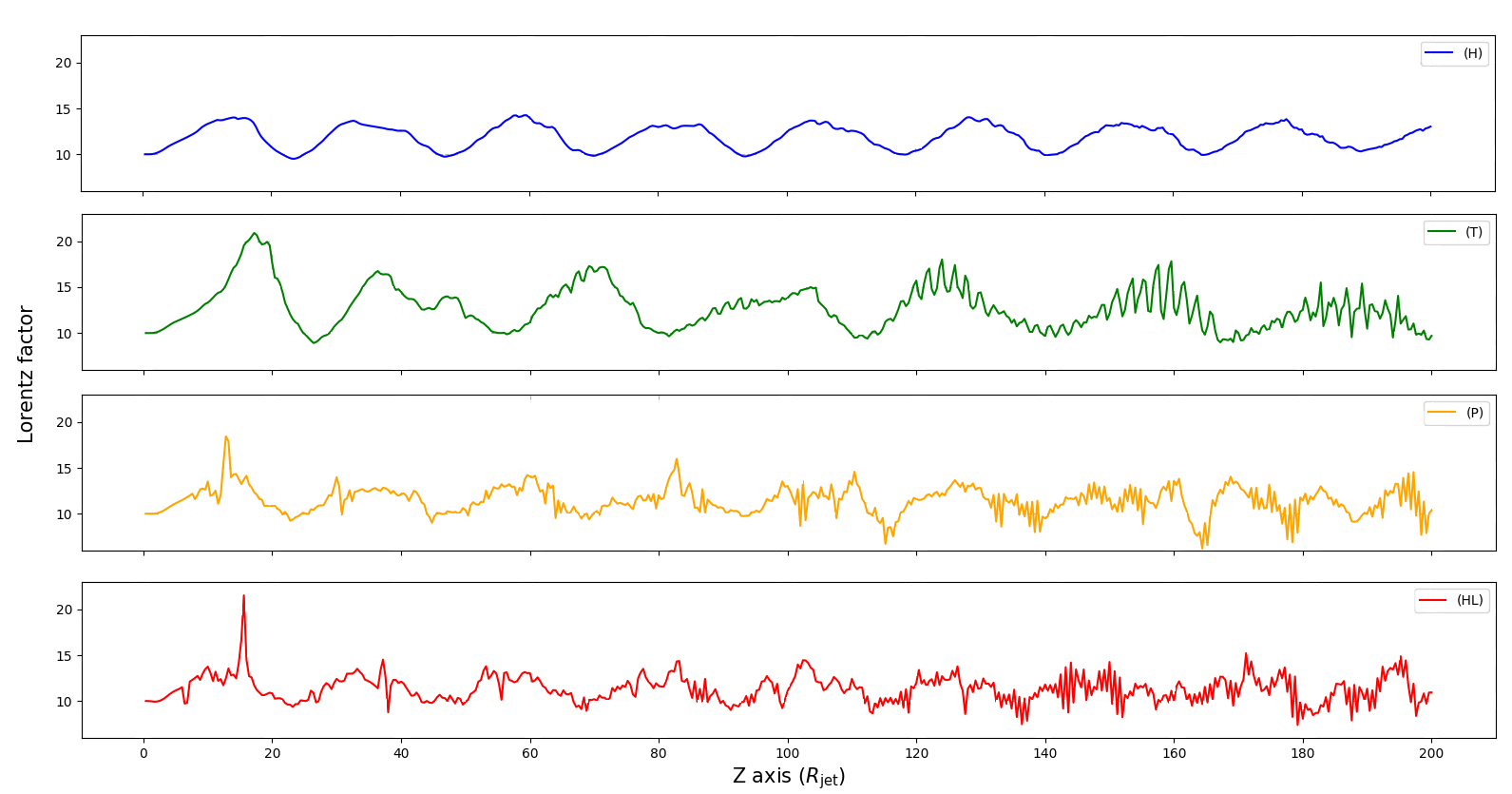}
    \caption{Profile of the mean Lorentz factor along the propagation axis $Z$ and for a radius between $R = \left[0.1, 0.23 \right]~R_{\rm jet}$ (in the inner jet). We show the profiles of the 4 cases studied with (\textbf{H}) in blue, (\textbf{T}) in green, (\textbf{P}) in orange and (\textbf{HL}) in red without ejecta.}
    \label{fig: tri lfac}
\end{figure*}

\subsection{Hydrodynamic case (\textbf{H})}
\label{subsect: jet H}

The over-pressured inner jet expands inducing the development of multiple steady conical recollimation shocks and rarefaction waves along the inner and the outer jet component (Fig.~\ref{fig: 2D contour energy, Lorentz factor - H T P HL}, top left). The high inertia ratio between the inner and outer jet component $\left(\gamma^2_{\rm in} h_{\rm in} \rho_{\rm in}\right)/\left(\gamma^2_{\rm out} h_{\rm out} \rho_{\rm out}\right)\simeq 42$ enhances the influence of the inner component on the outer component, even if the inner component carries only $25\%$ of the total kinetic energy flux. \\
In the jet inlet at the inner/outer jet component interface, the lateral expansion of the over-pressured inner jet is associated with the development of conical rarefaction waves propagating in the inner jet component toward the jet axis, and conical shock waves propagating toward the jet edge in the outer jet. These propagating waves form an angle $\alpha =\arctan\left( 1/{\cal M}\right)$ with the jet axis, where ${\cal M}=\gamma\,v/(\gamma_{\rm s} c_{\rm s})$ is the relativistic Mach number of the jet component in which the waves propagate, $c_{\rm s}$ is the sound speed and the associated Lorentz factor $\gamma_{\rm s}=1/\sqrt{1-c_{\rm s}^2}$ (for more details see \citep{Hervet_2017}). These waves produce a pattern of successive and near-equidistant standing recollimation shocks with separation distance of $\delta Z_{\rm shock}=2 R_{\rm jet} {\cal M}$. \\
In the inner jet component, the rarefaction wave propagates inward and forms an angle $\alpha_{\rm in} \sim 5.5 \degree$ with the jet axis; when it reaches the jet axis, it is reflected outward under the same angle. At the interface between the inner and outer jet components, the wave is partially transmitted as a rarefaction wave in the outer jet with an angle  $\alpha_{\rm out} =\arctan\left( 1/{\cal M_{\rm out}}\right) \simeq 7.5\degree$ and it is partially reflected as a shock wave toward the jet axis inside the inner jet.\\
Each time the wave is partially reflected at the inner or the outer jet border, it changes the type from rarefaction to shock wave and vice versa. We can clearly see this structure in the evolution of the inner-jet Lorentz factor along the $Z$-direction (Fig. \ref{fig: tri lfac}, blue curve). The partial  transmission from the inner toward the outer jet dampens the wave and its intensity decreases with distance, whereas the wave intensity associated with the outer jet component increases with distance, since it accumulates the successive waves transmitted from the inner component (Fig. \ref{fig: 2D contour energy, Lorentz factor - H T P HL}). Moreover, each time the waves from inner and outer jet interact, a partial wave reflection is produced toward the jet axis. A pattern of multiple waves arises with a wavelength depending on the Mach number of the inner and outer jet and on the jet radius. \\
The transverse ram pressure produced by this reflected wave causes an expansion of the jet. The expansion of the inner jet component with its higher effective inertia is more pronounced. The radial expansion stabilizes fairly quickly and the jet opening angle becomes $\theta_{\rm jet} \simeq 0.05 \degree$ for the parameters chosen in our simulations (Fig. \ref{fig: theta tri}, blue line). \\

\subsection{Toroidal case (\textbf{T})}
\label{subsection : jet T}

\begin{figure}[h]
    \centering
    \includegraphics[width=\columnwidth]{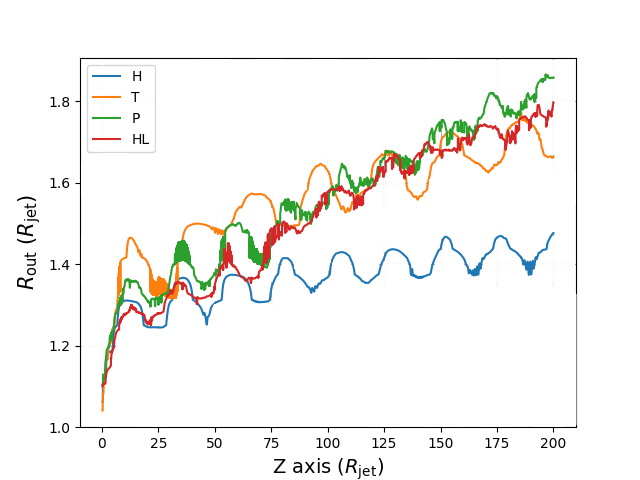}
    \caption{Radius of the external jet ($R_{\rm out}$) as a function of the distance along the jet axis. We represent, in different colors, the hydrodynamic (\textbf{H}, lowest curve), the toroidal (\textbf{T}), the poloidal (\textbf{P}) and the helical case (\textbf{HL}) of jet. The opening angle is deduced from the slope of a linear function fitted to these curves.}
    \label{fig: theta tri}
\end{figure}

\begin{figure}[h]
    \centering
    \includegraphics[width=\columnwidth]{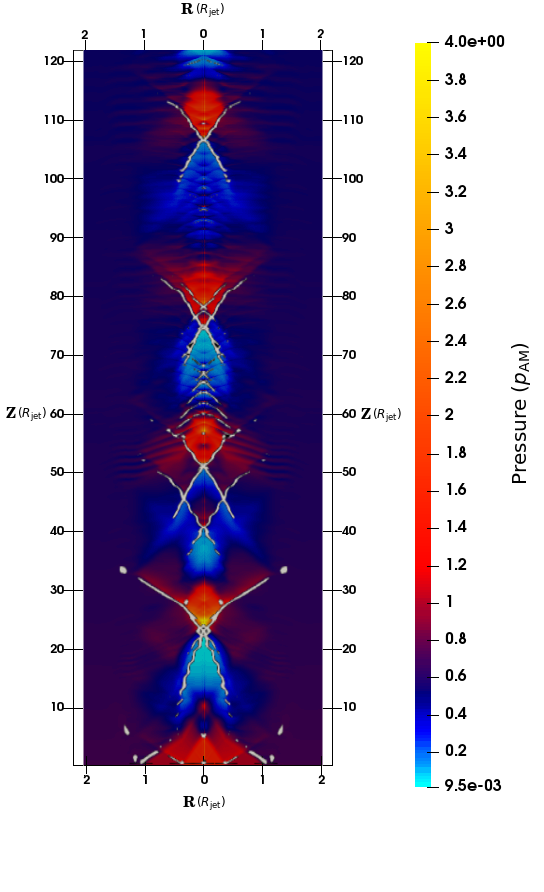}
    \caption{Zoom on the base of the toroidal jet (\textbf{T}). A standing shock structure appears on the pressure map.The $R$-axis and the $Z$-axis are given in $R_{\rm jet}$ unit. The white lines represent the rarefaction and compression wave shock fronts.}
    \label{fig: zoom on shock}
\end{figure}

In the toroidal case, a large-scale azimuthal magnetic field is set up in the inner jet and in the outer jet with respectively $B_{\rm in, \varphi} = 50$ mG and $B_{\rm out,\varphi} = 5$ mG. Due to the over-pressured inner jet, a recollimation structure arises in the jet, with a "diamond" structure formed by compression and rarefaction waves in the two components (Fig. \ref{fig: 2D contour energy, Lorentz factor - H T P HL} (top right) or Fig. \ref{fig: zoom on shock}). Since the magnetic field strength chosen for this case is weak, the kinetic energy flux and the inertia ratio between the outer and the inner jet component remain of the same order as for the hydrodynamic case (\textbf{H}). 
However, there are a few differences between the toroidal and hydrodynamic cases. 

The high Lorentz factor associated with the inner jet component of at least $\gamma=10$ induces a strong radial electric field $E_{r}=-B_{\varphi}\cdot\sqrt{1-1/\gamma^2}$ that decreases the efficiency of the radial magnetic tension to collimate the jet. This efficiency decreases more in the rarefaction region where the Lorentz factor can reach a higher value of $\gamma=20$ and the magnetic field strength decreases. As a result, the jet expands radially in these zones and the rarefaction zones become more elongated in the Z direction. The shock wave at the recollimation knots is dampened and appears closer to the jet axis compared to the hydrodynamic case.
Therefore, the stationary shock wave decollimates the jet which expands with an opening angle two times higher than the hydrodynamic case with $\theta_{\rm jet} \simeq 0.10 \degree$ (Fig. \ref{fig: theta tri}, orange curve). Nevertheless, thanks to the magnetic tension, we recover a higher value of magnetic energy ($\sim B^2$) in the inner jet compared to an axial magnetic field. 
With distance from the jet inlet ($Z>100~R_{\rm jet}$), the strength of the shock wave and the associated rarefaction wave decreases. As a result, the ram-pressure applied by these waves weakens and the radial expansion of the jet slows down (Fig. \ref{fig: theta tri}, green curve). In this region, a radial instability grows in the jet inducing oscillations in the density and Lorentz factor.

\subsection{Poloidal case (\textbf{P})}
\label{subsection: jet P}

Now we consider the case of an initial large-scale axial and uniform magnetic field in the inner-jet  $B_{\rm in,~p}\,=\,50\, {\rm mG}$ and in the outer-jet $B_{\rm out,~p}\,=\,5\, {\rm mG}$.
As in the hydrodynamic case, the inner-jet component is over-pressured, which leads to multiple steady conical recollimation shocks and rarefactions waves that emerge at the jet inlet and propagate along the jet (Fig. \ref{fig: 2D contour energy, Lorentz factor - H T P HL}, top left).  
At the jet inlet, the low magnetization induces only a weak difference with the hydrodynamic case. The distance between two recollimation shocks remains of the same order. The difference appears at a larger distance, where the magnetized jet starts to decollimate. \\
When the shock waves interact with the inner and the outer jet interface, the jet undergoes a weak transverse expansion. In this case, the jet expansion induced by the shock wave is stronger than in the hydrodynamics case, with a jet opening angle three-times larger $\theta_{\rm jet} \simeq 0.17 \degree$ (Fig. \ref{fig: theta tri}, orange curve). The successive shock waves push the magnetic field toward the axis, increasing the magnetic pressure. As a result, the jet decollimates. Moreover, the rarefaction regions exhibit a stronger radial expansion, inducing more efficient acceleration. \\
The interaction between the shock waves and the poloidal magnetic field induces radial instabilities that can be associated with the development of thin structures. By perturbing the jet, these instabilities develop themselves at the interface where the expansion is at a maximum and where the poloidal magnetic pressure in $Z$-direction increases. These instabilities become more pronounced with distance from the core and disturb the jet. As a result, the intensity of the shock and rarefaction waves decreases with distance in comparison with the hydrodynamic case. In addition, an expansion of the external jet toward the internal jet is well marked. This expansion, due to the heating of the external jet and the magnetic pressure along the $Z$-axis, tends to compress and recollimate the internal jet and thus modifies the structure of stationary shocks. It is this compression that tends to stretch the shock structure in the internal jet along the propagation axis.

\subsection{Helical case (\textbf{HL})}
\label{subsection: jet HL}

Finally, we consider a helical case with similar characteristics as the poloidal case, with a magnetic field strength of $B = 50$ mG in the inner jet and $B = 5$ mG for the outer jet. The only difference is the helical structure with a pitch angle between the azimuthal and axial magnetic direction fixed to  $45 \degree$. \\
Overall, the recollimation shock structure is similar to the poloidal one (Fig. \ref{fig: 2D contour energy, Lorentz factor - H T P HL}, bottom right). As for the poloidal case, the jet decollimation occurs after the second recollimation knot at a distance $Z \simeq 50~R_{\rm jet}$ from the jet inlet, and the jet opening angle tends to $\theta_{\rm jet} \simeq 0.17 \degree$ (Fig. \ref{fig: theta tri}, red curve). A small difference in the jet expansion appears between the poloidal and helical case at a large distance that results from the toroidal magnetic tension that tends to collimate the jet. \\
As in the poloidal case, radial instabilities develop due to the axial magnetic pressure increasing in the rarefaction region and they explain the strong variation of the Lorentz factor along the $Z$-direction in the inner jet (Fig. \ref{fig: tri lfac}, red curve). \\
We observe again an expansion of the external jet toward the internal jet. The poloidal component of the magnetic field still involves compression of the internal jet and an elongation of the stationary shock structure in the direction of propagation. In addition and similar to the toroidal case, the toroidal component of the magnetic field implies a higher value of magnetic energy (compared to the case without toroidal component) in the inner jet. 

\section{Results from simulations of moving ejecta}
\label{sec : ejecta}

A promising scenario to explain the observed flux variability in AGN jets is to consider the propagation of shock waves within the jet. In our models, the shock wave is caused by an initially over-dense ($\rho_{\rm e} = 10^{3}\rho_{\rm in}$) spherical ejecta with radius $R_{\rm e} = R_{\rm jet}/6$ (half of the inner jet radius) and placed initially at the jet axis ($R = 0$) at the distance $Z = R_{\rm e}$ from the inner boundary. Its Lorentz factor is the same as the one of the inner jet $\gamma_{\rm e} = \gamma_{\rm in}$. With this configuration, the kinetic energy flux associated with the ejecta is $10^{47}~\text{erg/s}$. We should note that the thermal energy of the ejecta is negligible in comparison to kinetic energy. All the time in this section is given in the co-mobile frame ($\delta = 1$). \\
The ejecta reaches the edge of the simulation box at time $t = 200~R_{\rm jet}$, but the simulations run until $t = 230~R_{\rm jet}$ to cover the jet relaxation phase after the passage of the shock wave induced by the ejecta. The jet is presented with a moving shock wave in Fig. \ref{fig: 2D contour energy, Lorentz factor - H T P HL ejecta} corresponding respectively to the co-moving times $135~R_{\rm jet}$. 

\subsection{Hydrodynamic case (\textbf{H})}
\label{subsect: H ej}

For the hydrodynamic case (\textbf{H}) (Fig.~\ref{fig: 2D contour energy, Lorentz factor - H T P HL ejecta}, top left), the ejecta remains well confined within the inner jet during a short period $t<5~R_{\rm jet}$ until it starts interacting with the first standing shock. During this phase, the ejecta and thus the moving shock wave undergoes an adiabatic acceleration in the first rarefaction zone to reach a Lorentz factor $\gamma_{\rm ms}\simeq 14$ before it collides with the first standing shock. As a result, the thermal energy of the ejecta increases and as it evolves in the rarefaction region, it is accelerated once more (Fig. \ref{fig: lfac max ej}, blue curve). In interactions between the ejecta and internal shocks, all the gained energy of the ejecta is transferred to shock waves that continue to propagate mainly within the inner jet with low transverse energy loss. Globally, the velocity of the ejecta will follow the profile of the Lorentz factor of the internal jet without seeing its velocity drop to the minimum value of $\gamma = 11$. 
As the ejecta is traveling through the jet, it will perturb momentarily the stationnary shock structure.

\subsection{Toroidal case (\textbf{T})}
\label{subsect: T ej}

In the toroidal case (\textbf{T}) (Fig.~\ref{fig: 2D contour energy, Lorentz factor - H T P HL ejecta}, top right),  the interaction of the moving shock wave with the standing shocks has some similarities with the previous case. In the first rarefaction zone, the ejecta accelerates adiabatically before it starts interacting with the first shock. The resulting moving shock wave is subject to a strong radial expansion after this first interaction and slows down close to its initial value of $\gamma_{\rm ms} = 10$ at $Z \sim 50~R_{\rm jet}$. Then the moving shock sees its velocity increase between each recollimation shock (Fig. \ref{fig: lfac max ej}, green curve). 
Due to its initial higher inertia compared to the surrounding jet, the ejecta undergoes a stronger interaction when its crosses the recollimation zones. Therefore, the thermal pressure of the ejecta increases more than the ambient flow. Afterwards, in the rarefaction zone, the higher pressure and inertia of the shocked ejecta starts to behave as a fireball within the surrounding jet. As a result, the moving shock wave induced by the ejecta reaches a higher Lorentz factor than the surrounding jet.

As mentioned before (Section \ref{subsection : jet T}), the rarefaction zones are larger than in the other cases, especially after the fourth standing shock, where the shock wave is strongly accelerated to a value of $\gamma_{\rm ms} \simeq 30$ at $Z \sim 125~R_{\rm jet}$. Then the moving shock interacts strongly with all the following standing shocks and causes them to oscillate along the jet axis (with a typical oscillation time close to $\sim 13 R_{\rm jet}$).

\begin{sidewaysfigure*}
\centering

\begin{multicols}{2}

    \includegraphics[width = \columnwidth]{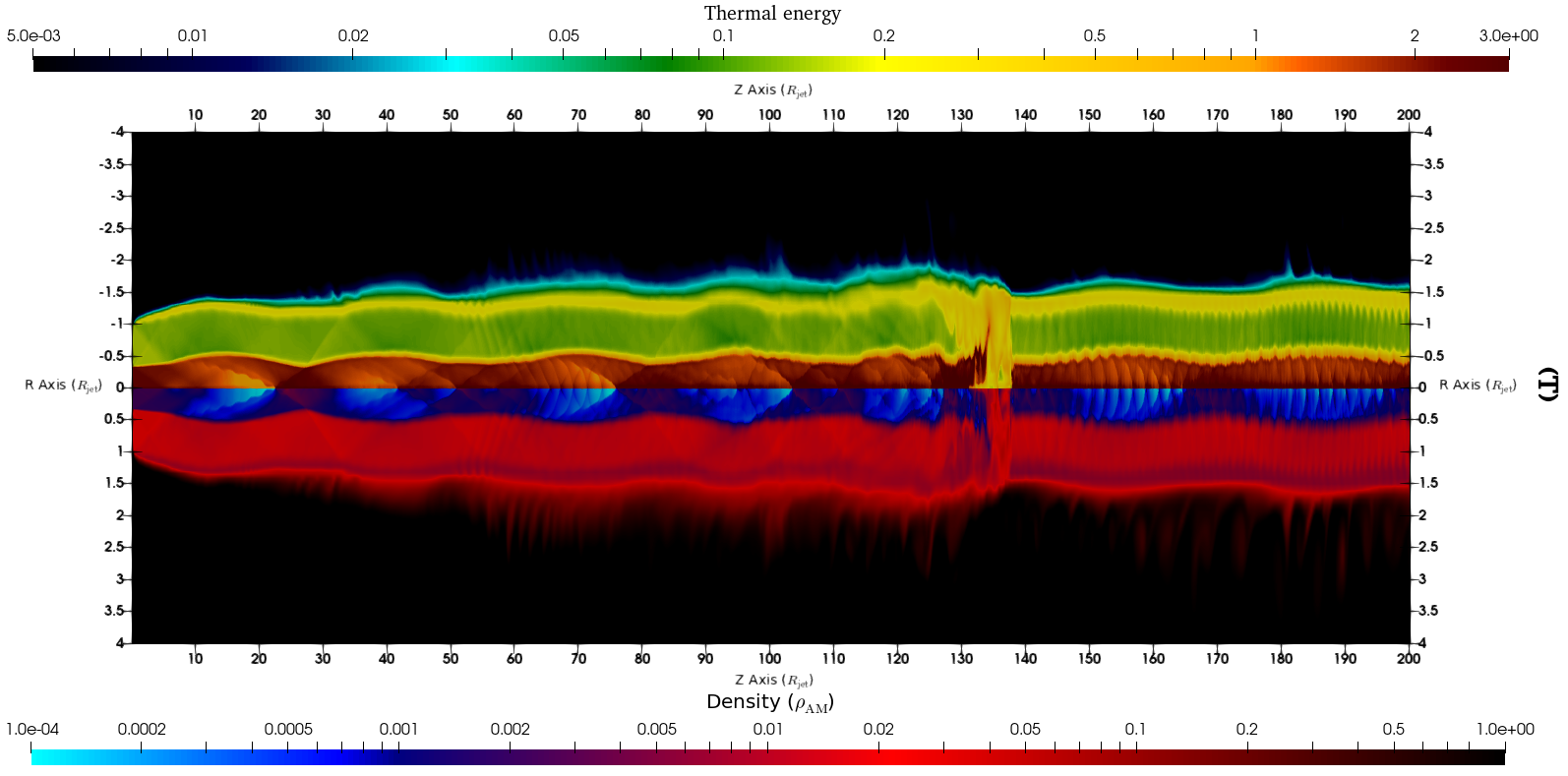}

    \includegraphics[width = \columnwidth]{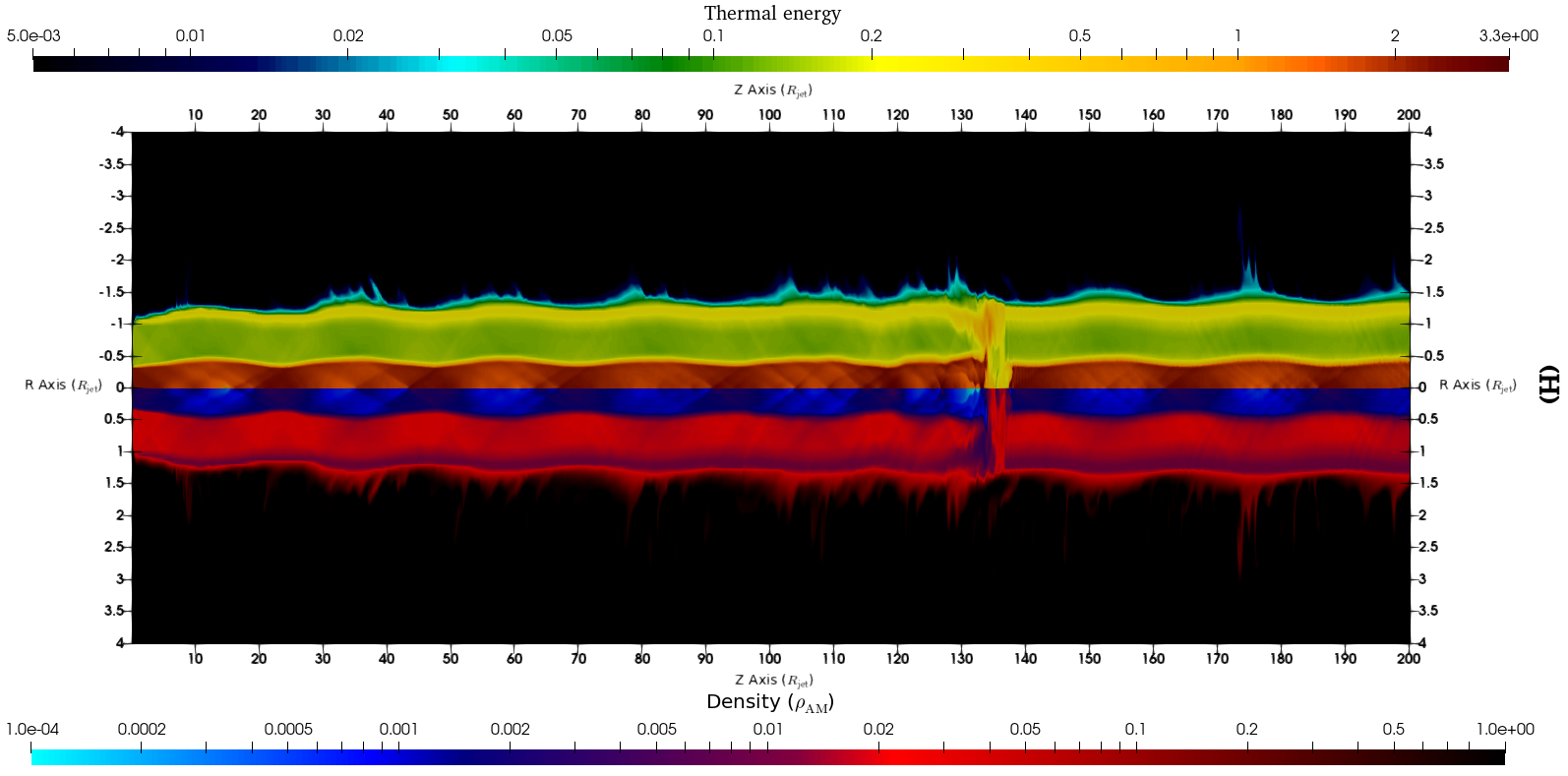}
    
\end{multicols}
\begin{multicols}{2}

    \includegraphics[width = \columnwidth]{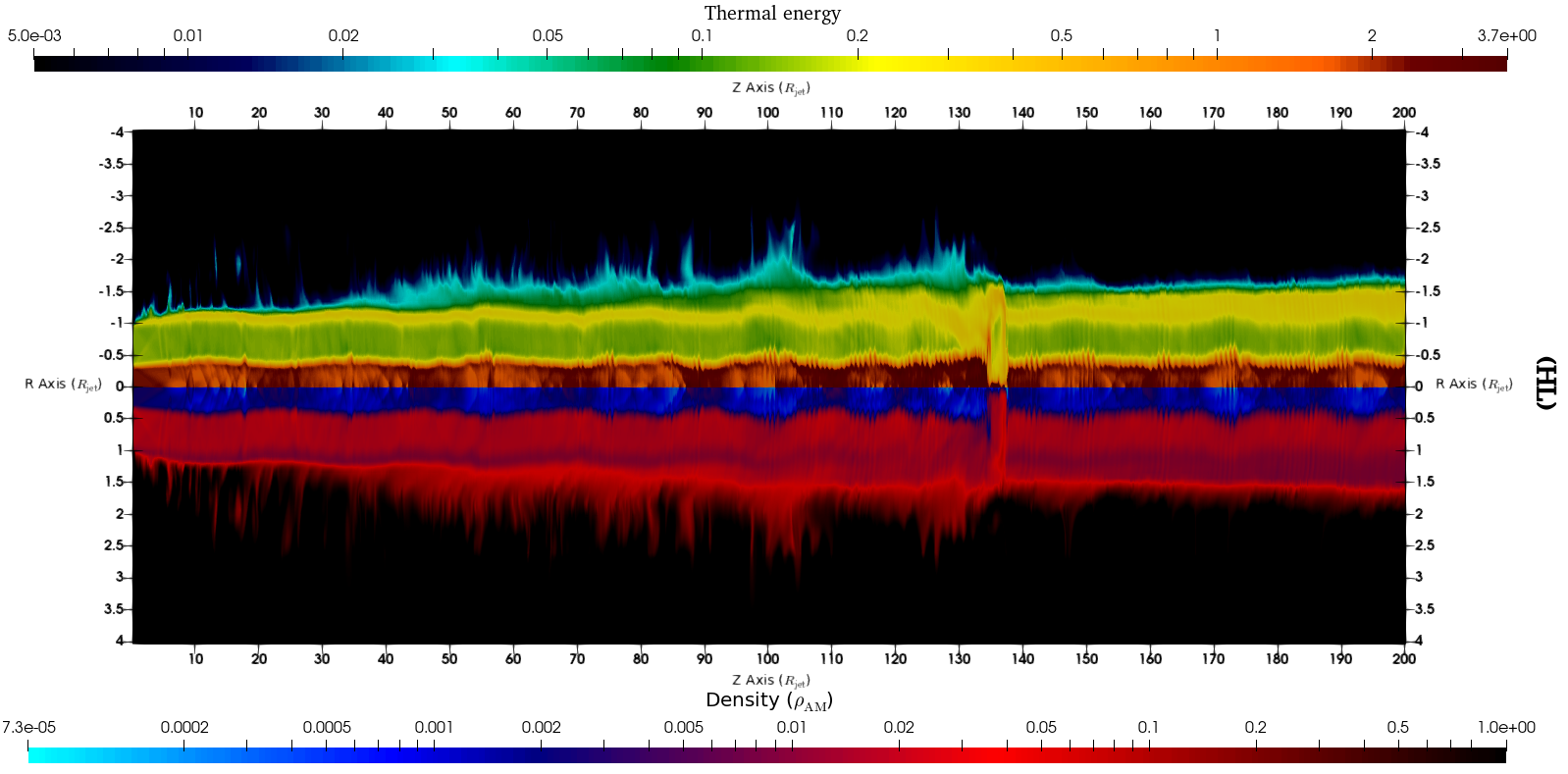}

    \includegraphics[width = \columnwidth]{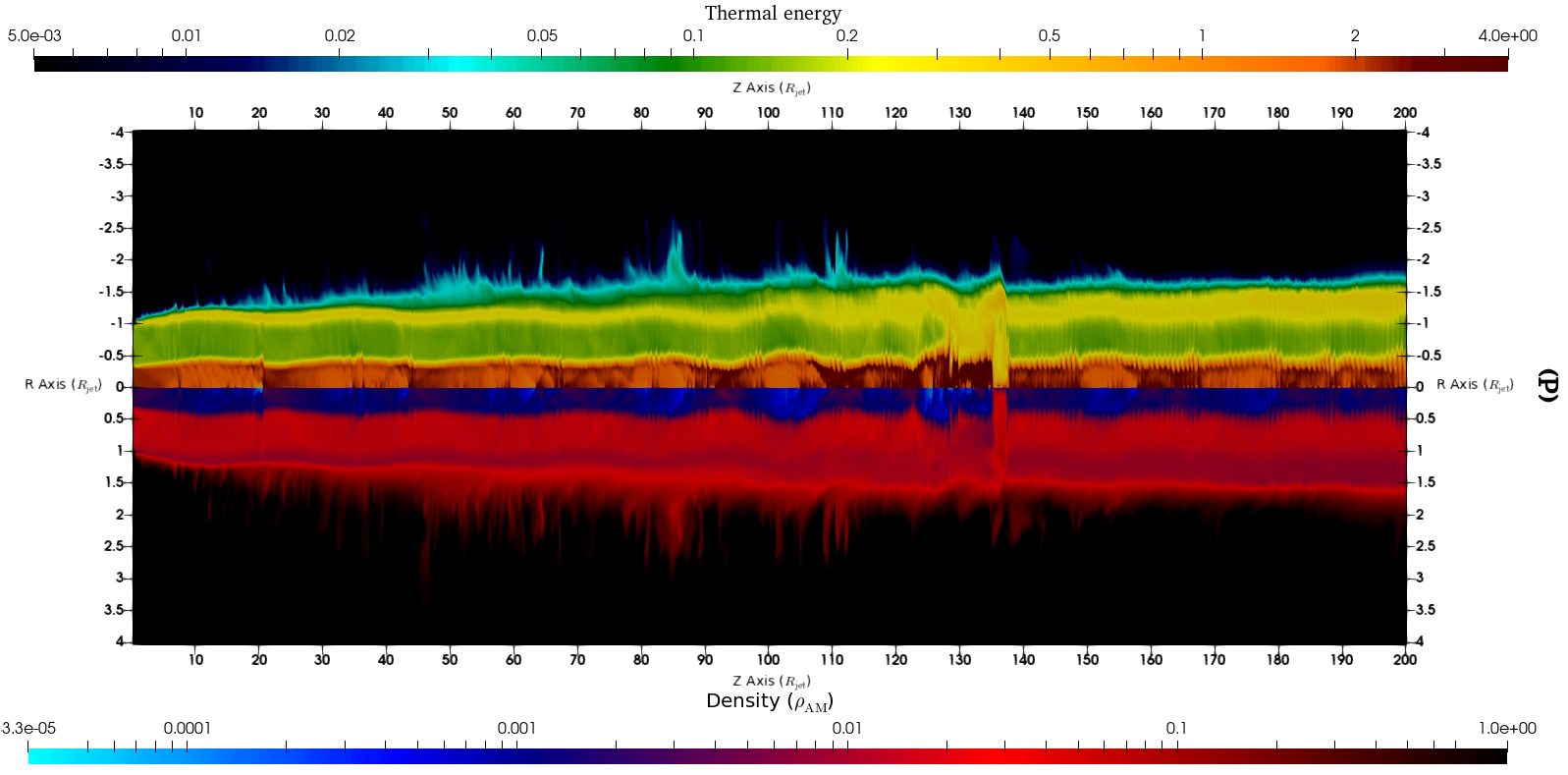}
    
\end{multicols}

\caption{Snapshot: different types of structured jets (\textbf{H}, \textbf{T}, \textbf{P} and \textbf{HL}) with an injected perturbation. The generated moving shock wave is located at $\sim 135 R_{\rm jet}$ from the base. The propagation of the jet is going from left to right along the increasing value of Z. For each case, the density contour (in log-scale) is drawn on the bottom side and the thermal energy contour on the top side. Units on x and y-axis in $R_{\rm jet}$ unit.}
\label{fig: 2D contour energy, Lorentz factor - H T P HL ejecta}

\end{sidewaysfigure*}

\begin{figure}[h]
    \centering
    \includegraphics[width=\columnwidth]{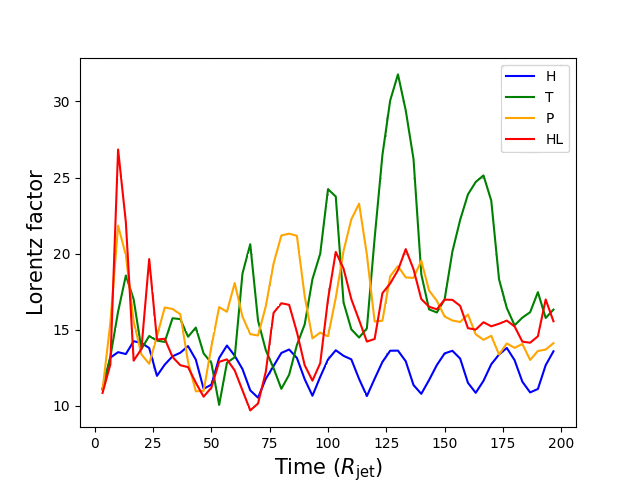}
    \caption{Evolution of the Lorentz factor of the moving shock as a function of time in the co-moving frame. We represent, in different colors, the hydrodynamic (\textbf{H}), toroidal (\textbf{T}), poloidal (\textbf{P}) and the helical case (\textbf{HL}) of jet.}
    \label{fig: lfac max ej}
\end{figure}

\subsection{Poloidal case (\textbf{P})}
\label{subsect: P ej}

In the poloidal case (\textbf{P}) (Fig.~\ref{fig: 2D contour energy, Lorentz factor - H T P HL ejecta}, bottom left), the moving shock wave undergoes a strong acceleration to reach  $\gamma_{\rm ms} \simeq 21$ before it interacts with the first stationary features and slows down to $\gamma_{\rm ms} \simeq 13$ at $Z \sim 24~R_{\rm jet}$ (Fig. \ref{fig: lfac max ej}, orange curve.) 
In a second phase, the resulting moving shock propagates through the successive rarefaction zones where it accelerates and the compression zones where it decelerates. The mean Lorentz factor of the moving shock increases with distance until it reaches a stationary shock at the distance $Z\sim 125~R_{\rm jet} $. This acceleration results from the expansion of the inner jet component in this region. Beyond, the moving shock continues the propagation in the inner jet component which is compressed by the outer jet, and transverse instabilities grow along the stationary shock wave. In this region, the rarefaction zones are smaller and they are subject to multiple small scale stationary shocks. As a result, the moving shock wave decelerates to reach a Lorentz factor of $14$.
As in the toroidal case, beyond the fourth stationary shock, the passage of the moving shock induces an oscillation.

\subsection{Helical case (\textbf{HL})}
\label{subsect: HL ej}

In the helical case (\textbf{HL}) (Fig.~\ref{fig: 2D contour energy, Lorentz factor - H T P HL ejecta}, bottom right), the moving shock, after crossing the first internal shock wave at $z= 5~R_{\rm jet}$, 
also undergoes a strong acceleration in the first rarefaction.
Like in the poloidal case, this strong acceleration is followed by a strong interaction of the moving shock with stationary shocks, leading to the deceleration of the moving shock to ($\gamma_{\rm ms} = 10$) (Fig. \ref{fig: lfac max ej}, red curve.) Beyond the fourth standing shock, the moving shock accelerates again to $\gamma_{\rm ms} \simeq 20$. 

\section{Modeling the radiative transfer}
\label{sec:radiative}

\subsection{Radiative processes}\label{subSec: Radiative processes}

In a post-processing step using a separate code, we evaluate the synchrotron radiation of an electron population in the radio band and solve the radiative transfer equation along a given line of sight. We construct synchrotron emission maps to study the variation of the flux observed in different zones along the jet over time, as well as light curves of the spectral flux density integrated over the full simulated jet. 
In each cell, the relativistic electrons population is set with a power law, as expected for shock acceleration,

\begin{equation}\label{Eq:radiation_Ne}
    N_{\rm e} \left(\gamma\right) \text{d} \gamma = K \gamma^{-\text{p}}~\text{d}\gamma\,,
\end{equation}
where $\gamma_{\rm min} < \gamma < \gamma_{\rm max}$. 
\noindent
Following \citep{Gomez_1995}, we define the normalization coefficient as,

\begin{equation}\label{Eq:radiation_coef_K}
    K = \left[ \dfrac{e_{\rm th,e} \left( \text{p} - 2 \right)}{1 - C_{\rm E}^{2-\text{p}}}\right]^{~\text{p}-1} ~ \left[\dfrac{1-C_{\rm E}^{1-\text{p}}}{n_{\rm e} \left( \text{p} - 1 \right)}\right]^{~\text{p}-2}\,,
\end{equation}
\noindent
where $e_{\rm th,e} = \epsilon_{\rm e} e_{\rm th}$ is the fraction of thermal energy carried by the electrons, $n_{\rm e} = \epsilon_{\rm e}\,n$ is the fraction of the electron number density with $\epsilon_{\rm e} = 0.1$, $\text{p} = 2.2$ is the index of the power law and the coefficient $C_{\rm E} = \gamma_{\rm max}/\gamma_{\rm min}$ is set to $10^3$. As a simplification, we do not take into account radiative losses of the relativistic electrons and assume, 

\begin{equation}\label{Eq:radiation_gmin}
    \gamma_{\rm min} = \dfrac{e_{\rm th,e}}{n_{\rm e}} \dfrac{\text{p}-2}{\text{p}-1} \dfrac{1 - C_{\rm E}^{1-\text{p}}}{1-C_{\rm E}^{2-\text{p}}} \,.
\end{equation}
In the present application, we are focusing only on the radio band, where the effect of radiative cooling is the smallest.\\

\begin{figure}[!h]
    \centering
    \includegraphics[width =\columnwidth]{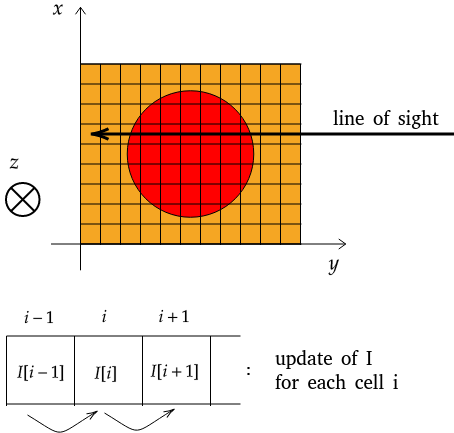}
    \caption{Schematic representation of the resolution of the radiative transfer equation. We sum the different contributions along the line of sight.}
    \label{fig:calcul_I}
\end{figure}

The specific intensity for each cell with index "\texttt{i}" is determined in the frame of a stationary observer at the location of the source (quantities in the co-moving frame are noted $x'$), 

\begin{equation}\label{Eq:radiation_Inu}
    I_{\nu; \,\rm \texttt{i}} = I_{\nu; \, \rm \texttt{i}-1} \exp{\left(-\tau_{\nu}\right)} + S_{\nu; \, \rm \texttt{i}} \left(1 - \exp{\left(-\tau_{\nu}\right)} \right)\,,
\end{equation}
where $\tau_{\nu}$ is the optical depth due to synchrotron self-absorption and $S_{\nu}$ is the  synchrotron source function.\\
Fig. \ref{fig:calcul_I} shows schematically how the contribution of each cell along the line of sight to the total specific intensity $I_{\nu}$ is estimated. It should be noted that here we do not account for light crossing time effects, which can lead to a superposition of the signal from different emission regions along the jet due to the relativistic movement of the jet and the final speed of signal propagation. In the radio band, this effect is expected to be of minor importance \citep{Chiaberge_1999}. 

The specific intensity $I_{\nu}$ $\left(~\text{in}~ \left[\text{erg}~\text{s}^{-1}~\text{cm}^{-2}~\text{Hz}^{-1}~\text{sterad}^{-1}\right]\right)$ depends on the specific emission coefficient $j_{\nu}$, the absorption coefficient $\alpha_{\nu}$ and optical depth $\tau_{\nu}$. They are transformed to the observer frame at the location of the source for each cell,

\begin{align}
    \label{eq : emissivité}
    j_{\nu} &= \delta^2~j_{\nu'}\,, \\
    \label{eq: absorption}
    \alpha_{\nu} &= \delta^{-1}~\alpha_{\nu'}\,, \\
    \label{eq: profondeur optique}
    \tau_{\nu} &= \tau_{\nu'}\,.
\end{align}

These transformations \citep{Rybicki_1979} depend on the Doppler factor $\delta = \left( \gamma \left( 1 - \beta \cos{\left( \theta_{\rm obs} \right)} \right) \right)^{-1}$ with $\gamma=1/\sqrt{1-v^2}$ the bulk Lorentz factor of the material in the cell and $\theta_{\rm obs}$ is the angle between the direction of the jet axis and the line of sight. The different quantities ($j_\nu$, $\alpha_\nu$ and $\tau_\nu$) are estimated in each cell following the approximations given by \citep{Katarzynski_2001} which are appropriate for the cases studied here. 
The synchrotron flux in the observer frame on Earth is determined by,

\begin{equation}\label{eq: rad_fluxFnu}
    F_{\nu} = \frac{S_{\rm e}}{d_{l}^2} \left( 1 + z \right) I_{\nu}\,,
\end{equation}
with $d_{\rm l}$ the luminosity distance (assuming $H_0 = 70~\text{km.s}^{-1}\text{Mpc}^{-1}$) and $S_{\rm e}$ the typical emission surface.\\
In our study, as an illustration, we choose the distance corresponding to M\,87 ($z = 0.00428$). Finally, we obtain a 2D flux map of the synchrotron emission. To provide images that can be eventually compared with real VLBI images, we smooth the simulated images with a typical beamwidth obtained in the radio domain for M\,87 close to $1.6~R_{\rm jet}$. To be able to distinguish between the emission from the jet and from the ejecta, we adjust an asymmetric 2D Gaussian distribution to the ejecta at each time step and extract the flux from the fitted ($2\sigma$) region (Fig. \ref{fig: CL results 90}).\\
To add synchrotron emission in the hydrodynamical case (\textbf{H}), a passive turbulent magnetic field is added during the post-processing step. In this case, we assume that the magnetic energy density is a fraction $\epsilon_{\rm B}$ of the thermal energy density,

\begin{equation}
    B_{\text{turb}} = \sqrt{\epsilon_{\rm B}~\epsilon_{\rm the}}\,,
    \label{eq: Bturb}
\end{equation}
with $\epsilon_{\rm B} = 0.1$. 

\subsection{Synthetic synchrotron images and light curves}
\label{sec: radiative transfer results}

\begin{figure*}
    \centering
    \includegraphics[width=2\columnwidth]{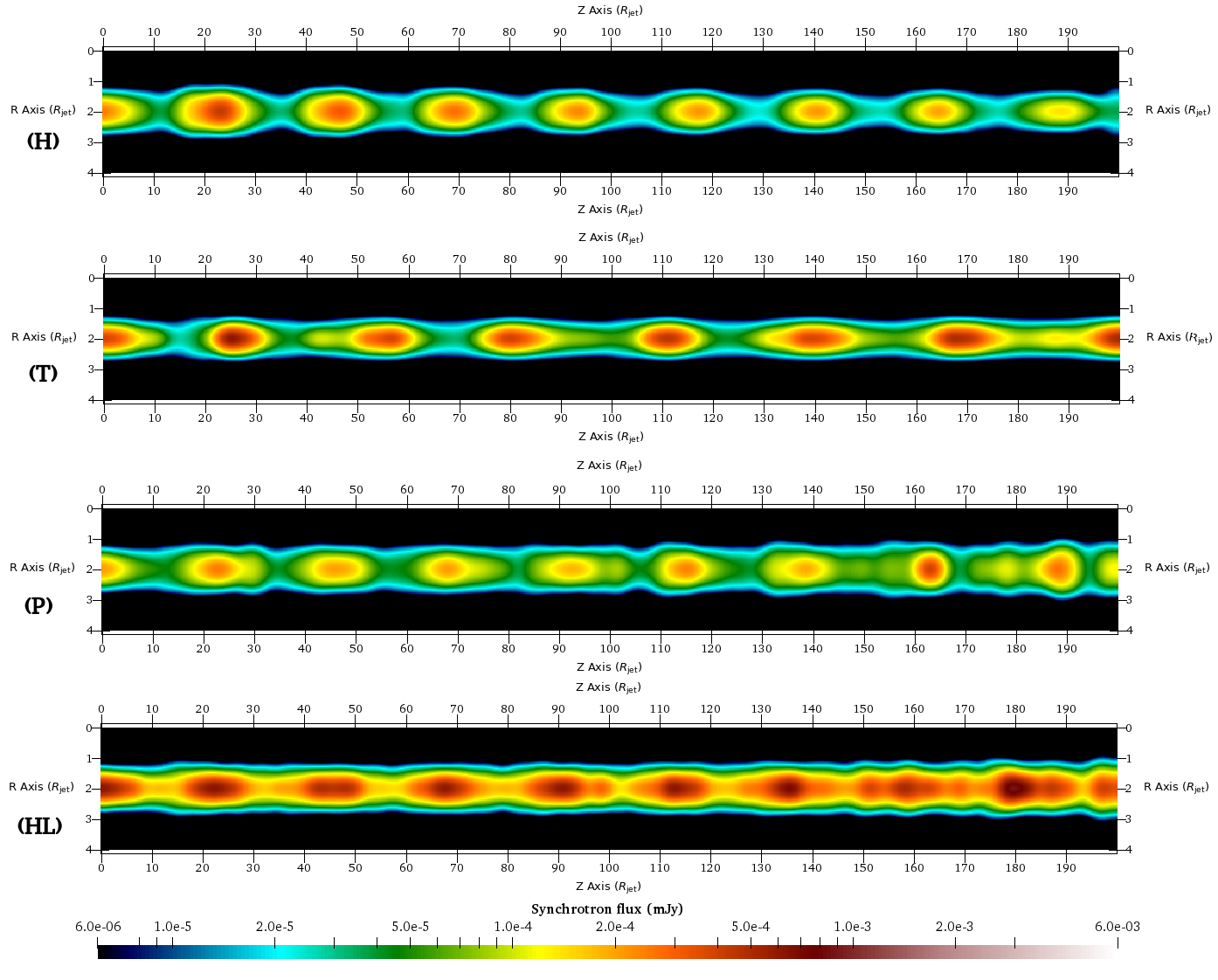}

    \caption{Snapshot : synchrotron emission map of the different types of jets (\textbf{H}, \textbf{T}, \textbf{P}, and \textbf{HL}) without ejecta and stationnary. Each map represent the flux intensity in mJy unit. The $R$ and $Z$-axis are given in $R_{\rm jet}$ unit. These maps represent a resolution of the radiative transfer equation with an angle between the jet propagation axis and the line of sight equal to $\theta_{\text{obs}} = 90 \degree$ and $\nu = 1$ GHz.}
    \label{fig: PP results initally}
\end{figure*}

\begin{figure*}
\centering
    \includegraphics[width=2\columnwidth]{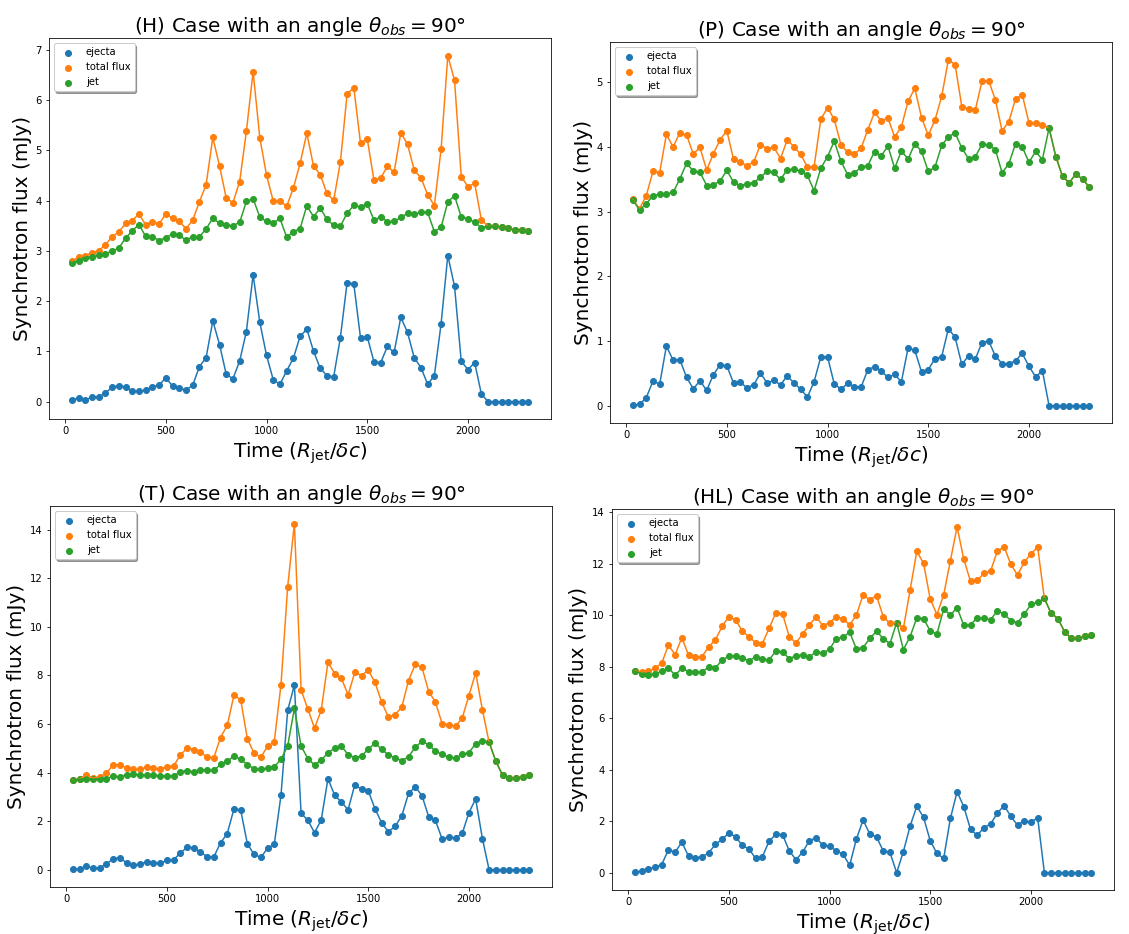}
    
    \caption{Light curve obtained by integrating the total synchrotron flux emitted from a simulation box with a size of $\left[ \text{R} = 8,~\text{Z} = 200 \right]R_{\rm jet}$. The computation of the light curve is realized from the four different cases of jets (\textbf{H}, \textbf{T}, \textbf{P} and \textbf{HL}). The flux is integrated from $\text{t} = 0$ the time of the injection of the ejecta, until $t \sim 2300~R_{\rm jet}/\delta$  with $\delta(\theta_{\rm obs} = 90 \degree) = 0.1$ in the absolute mobile frame and with $\nu = 1$ GHz. We separate the total flux (orange) in two component : the jet (green) and the moving shock wave (blue).}
    \label{fig: CL results 90}
    
\end{figure*}

\begin{figure}[h]
\centering

    \includegraphics[width=\columnwidth]{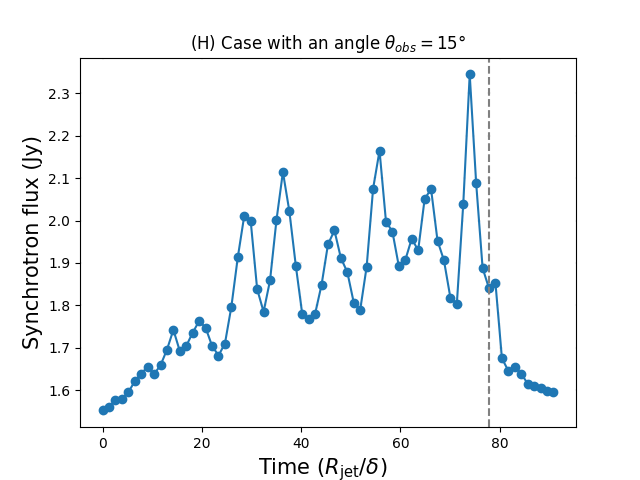}

    \includegraphics[width=\columnwidth]{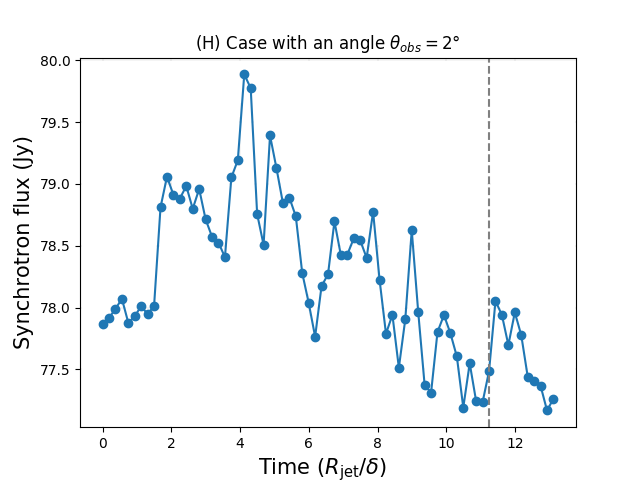}
    
    \caption{Light curve obtain by integrating the total synchrotron flux emitted from a simulation box with a size of $\left[ \text{R} = 8,~\text{Z} = 200 \right]R_{\rm jet}$. The computation of the light curve is realized for the hydrodynamic case (\textbf{H}) of jet. The flux is integrated from $\text{t} = 0$ the time of the injection of the ejecta, until respectively $t \sim \left(90,~13 \right)~R_{\rm jet}/\delta$ for $\delta(\theta_{\rm obs} = 15 \degree) \simeq 2.57$ (top) and $\delta(\theta_{\rm obs} = 2 \degree) \simeq 18$ (bottom) in the absolute frame and $\nu = 1$ GHz. The gray dotted line represent the exit of ejecta from the simulation box.}
    \label{fig: CL results HL}
\end{figure}

 Synthetic images are build by integrating the radiative transfer equation (Eq.~\ref{Eq:radiation_Inu}) along a line of sight, for all four cases (\textbf{H},\textbf{T},\textbf{P}, \textbf{HL}) and for one viewing angle $\theta_{\text{obs}}={ ~ 90\degree}$ (Fig.~\ref{fig: PP results initally}).
 We show the light curves realized for the four cases tested at $\theta_{\rm obs} ={~ 90 \degree}$ (Fig. \ref{fig: CL results 90}) and for the helical case at $\theta_{\rm obs} ={2\degree,~ 15\degree}$ (Fig. \ref{fig: CL results HL}).
The light curve is computed by integrating the flux on the full simulation box for the three observation angles. For an observation angle of $\theta_{\text{obs}}=90\degree$, it is also computed by integrating only the emission issued from the moving shock wave using the method described in Section~\ref{subSec: Radiative processes}. \\

The synthetic image of the stationary hydrodynamic jet (\textbf{H}) emission obtained for the observation angle $\theta_{\rm obs}=90\degree$ is presented in Fig.~\ref{fig: PP results initally} (top). The emission is dominated by the stationary knots within the inner jet component. The intensity and the size of the emitting knots decrease with distance as a result of the slow damping of shock strength with distance (Section~\ref{sec : the two-component model}). The contribution from the external jet component to the emission is negligible, since the stationary shock wave within this jet component is weak.

The interaction between the moving shock wave, induced by the ejecta, with the stationary shocks rapidly increases the injection of accelerated particles and  the associated synchrotron emission, causing a flare. The emission from the moving shock wave increases each time it passes through a standing knot. Moreover, after each collision, the stationary compression wave within the jet cannot ensure the radial cohesion of the shocked knot, thus the heated knot undergoes adiabatic expansion until it reaches the interface between the inner-outer jet. Afterwards the knot cools down and contracts. A remnant emission is associated to this adiabatic expansion. This illumination is visible on the jet light curve obtained with observation angle $\theta_{\rm obs}=90\degree$ (Fig. \ref{fig: CL results 90}, top left). It contributes to the emission during the decay phase of the observed flares (Fig.~\ref{fig: CL results 90}, top left).   
Moreover, each time a knot is shocked by the moving shock wave, it starts to stretch and to oscillate along the jet axis. This behavior is associated with local changes of jet characteristics, such as the Mach number $\cal{M}$ and the inner jet radius $R_{\rm jet, in}$. As a result, the structure of the standing shock waves evolves over a long period. This variability is associated with an increase of the base emission of the jet compared to the stationary case. \\

The synthetic image of the stationary state for the toroidal case (\textbf{T}) is shown in the second panel of Fig.~\ref{fig: PP results initally}. Like in the hydrodynamic case (\textbf{H}), the emission comes mainly from the standing shocks in the inner jet, but the emission from the region between the shocks is stronger. Moreover, the stationary shocks appear more elongated along the jet and especially beyond the fourth knot. This behavior results from the toroidal magnetic field (Section \ref{subsection : jet T}). 

The moving shock within the jet increases the emission, and the interaction with the stationary knots produces flares (Fig. \ref{fig: CL results 90}, bottom left). In comparison with the hydrodynamic case (\textbf{H}), the strength of the flares is variable with time. The strongest flare occurs at a time $t \sim 1100 R_{\rm jet}/\delta$ when the moving shock crosses the fourth (and strongest) standing shock. This interaction leads to a strong deformation and oscillation of the knots along the jet axis. The following relaxation of the knots is associated with a slow flux decline in the light curve and produces an asymmetric flare. This asymmetry occurs in the strongest flares and is enhanced when a knot splits in two after interacting strongly with the moving shock. 

The synthetic image for the poloidal case (\textbf{P}) is shown in the panel on  Fig.~\ref{fig: PP results initally}, where the emission of the stationary shock structure is represented. The emission comes mainly from the part of the stationary shocks very near to the jet axis, where the shock is sufficiently strong. In comparison to cases \textbf{H} and \textbf{T}, the steady shock wave is weaker in the poloidal case. A complex emission structure is visible due to the magnetic pressure along the $Z$-axis and the expansion of the outer toward the inner jet (Section \ref{subsection: jet P}). In this case, the variations in flux observed between shock and rarefaction zones are weak. Due to the expansion of the outer jet, the emission structure is more elongated along the jet axis and therefore less pronounced. 
 As before in the \textbf{P} case, the moving shock induced by the ejecta triggers flares as it crosses the steady knots (Fig. \ref{fig: CL results 90}, top right) but the efficiency of its interaction is weaker than in the \textbf{H} and \textbf{T} cases. Indeed, the poloidal magnetic field tends to damp the strength of the steady shock waves and it limits the transverse expansion of the moving shock wave. The resulting light curve is characterized by weak flares.
 Furthermore, when the moving shock interacts with a standing shock, instabilities form along the steady shock wave. These instabilities will propagate through the jet in the form of several moving shocks and thus locally increase slightly the synchrotron emission. 

Finally, the synthetic images of the helical case (\textbf{HL}) can be see in the last panel of Fig. \ref{fig: PP results initally}, where the emission of the stationary shock structure is represented. As in all the others cases, the emission comes mainly from the shock zones within the inner jet. However, the observed synchrotron flux is twice as high as in the other cases, due to the combination of the toroidal field effects with strong emission emanating from the center of the knots and the poloidal field effects with diffuse emission resulting from the compression of the inner jet by the outer component. \\
As in the previous cases, the moving shock will increase locally the synchrotron flux emitted and in particular when it crosses a stationary knot. But despite the fact that a toroidal magnetic field component is present, the shape of the light curve is very similar to the poloidal case (\textbf{P}) at $\theta_{\rm obs} = 90 \degree$ (Fig. \ref{fig: CL results 90}, bottom right). As the emission is extended along the axis in the internal jet, variability is observed essentially from the inner jet itself due to the propagation of the moving shock. The presence of a poloidal field allows the development of transverse instabilities along steady shock waves. These instabilities will propagate in the inner jet in the form of moving shocks behind the principal moving shock and increase the emission from the jet. We notice that the impact of a magnetic tension due to the toroidal component is visible when the moving shock interacts with the last shock zones. In fact, the moving shock interacts more strongly with the last standing knot where the most pronounced flare occurs. As before, the interaction of the moving shock with standing knot induces an knot oscillation along the jet axis. \\
 
To evaluate the impact of a reduced viewing angle on the shape of the light curve, we show the results for the hydrodynamic configuration (\textbf{H}) (Fig. \ref{fig: CL results HL}) for $\theta_{\rm obs}=15\degree$ and $\theta_{\rm obs}=2\degree$ for 1 GHz.
Compared to the one obtained for $\theta_{\rm obs}=90\degree$, they show the expected increase in the overall flux due to stronger Doppler beaming. The flares are shorter in duration due to the Doppler effect, but also due to self absorption. In certain cases, the dense ejecta partially hides the shocked knot behind it.
 For an observation angle $\theta_{\rm obs}=2\degree$, the self absorption effect is 
 very significant. The mean intensity decreases with distance, since, as the ejecta propagates within the jet, it hides a large number of stationary knots.  

\section{Comparison with radio observations, the case of 3C~273}
\label{sec : 3C273}

Our study shows the complex link between ejecta propagating in magnetized two-flow jets with stationary shocks, and its observed radio variability.
All simulations in the present study consider a kinetic power of the outer jet larger than the one of the inner jet with an initial ratio of 3:1. The ratio of two-flow kinetic powers was proposed in \cite{Hervet_2017} as a critical criterion discriminating between types of VLBI radio jets, which are themselves associated with spectral types of blazars \citep{Hervet_2016, Piner_2018, Lister_2019}. Two-flow jets with kinetic powers within the same order of magnitude, such as the ones simulated in this study, were found to be the most similar to FSRQ-type blazars (flat spectrum radio quasar). 

In order to compare the results of our simulations with an astrophysical case, we focus on the radio observations of one of the brightest and best monitored FSRQs over decades, 3C~273 (B1226+023). Its redshift of  $z = 0.1583$ \citep{Strauss_1992} translates into a scaling factor of 2.73 pc/mas considering $H_0 = 70$ km.s$^{-1}$.Mpc$^{-1}$.
3C~273 displays a peculiar mix of fast moving and quasi-stationary radio knots. This hybrid radio kinematic behavior is most often observed in intermediate blazars (Low or Intermediate frequency-peaked BL Lacs) \citep{Hervet_2016}.
However 3C~273 significantly differs from these sources in the way that quasi-stationary knots are visible only during low-activity periods of the jet, not continuously.

\begin{figure*}
\centering
    \includegraphics[width=2\columnwidth]{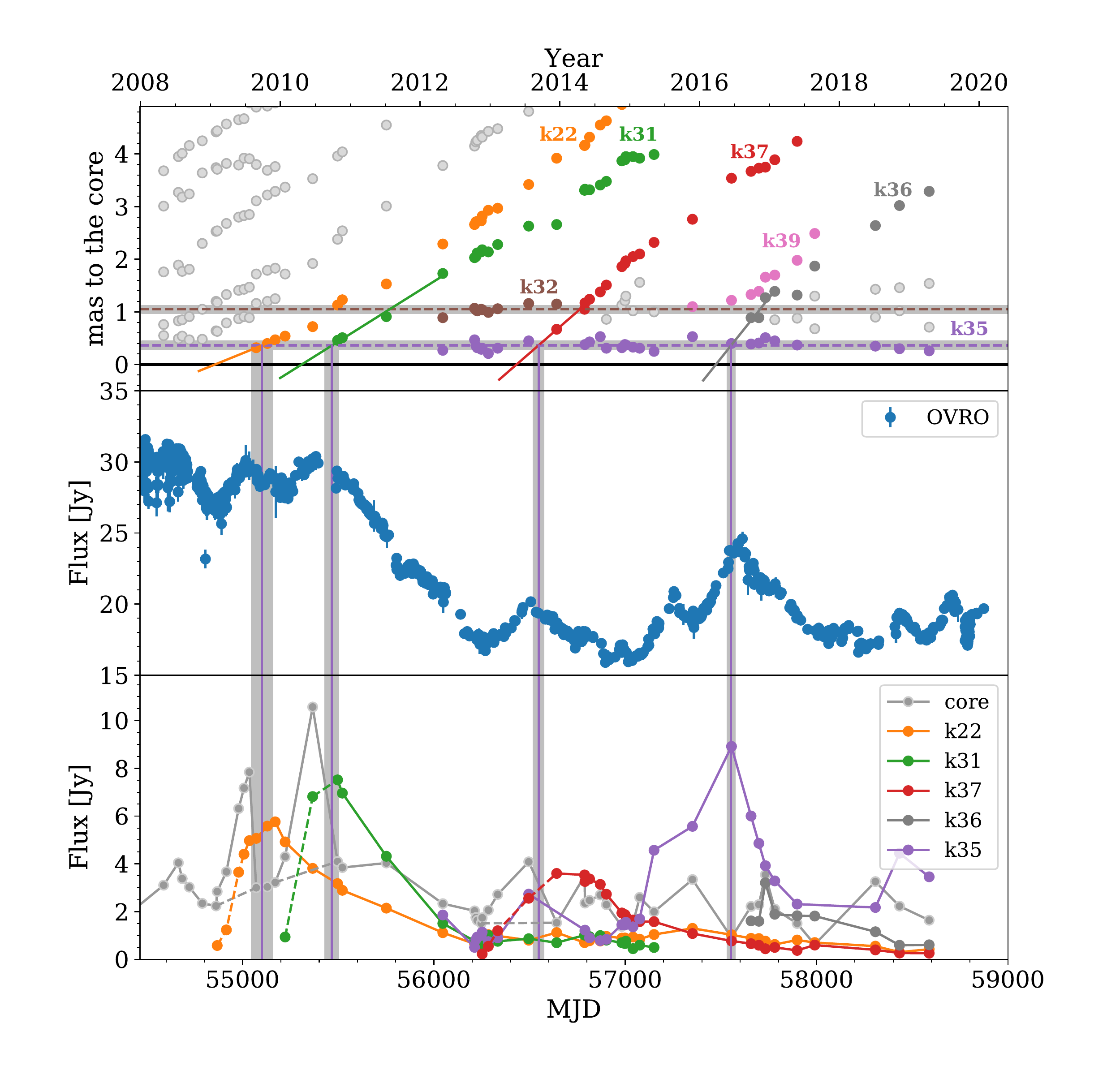}
    \caption{3C~273 observed at 15.3 GHz. \textit{Top panel:} Distance to the core of radio knots analyzed by MOJAVE. We focus on the firmly identified components (in color). Straight lines are linear extrapolations of the moving knots based on their first 4 observations. Horizontal dashed lines show the mean position in the jet of the two observed quasi-stationary knots k32 and k35, with the gray band displaying the 1 sigma dispersion around the mean. \textit{Middle panel:} radio jet light curve observed by OVRO. \textit{Bottom panel:} Measured flux of the radio core and moving knots. Dashed lines (extrapolation assuming smooth core emission) indicate that the observed core variability is actually due to the flux increase of the emerging moving knots when they are indiscernible from the core due to the limits of the VLBA angular resolution. Vertical lines show the most likely time when the moving knots pass through the stationary zone defined by k35 and the purple dashed line, with its associated uncertainty in gray.}
    \label{fig: 3C273}
\end{figure*}

For this study, we used 15.3 GHz observations from the VLBA, analyzed by the MOJAVE team up to August 2019 (Lister et al., in prep). Most of these data up to December 2016 are publicly available from \cite{Lister_2019}.
We combine this dataset with observations, at the same frequency, from the Californian 40m single-dish OVRO telescope, which provides public light curves from a monitoring program of Fermi-LAT blazars \citep{Richards_2011}. 
Our goal is to see, in a qualitative way, how the observed radio-VLBI ejecta influence the overall jet light curve observed with OVRO, as well as the luminosity evolution during their propagation. \\
We consider the period 2008-2019, overlapping both OVRO and MOJAVE observations with a dense VLBA monitoring. We specifically focus on four fast moving radio knots (k22, k31, k37, k36) and two quasi-stationary knots (k32, k35) which were observed during this period. All these observations are gathered in Figure \ref{fig: 3C273}. 
The moving knot k39 is not considered in our study as it is only referenced by MOJAVE beyond the k32 zone, and we suspect several wrong identifications between k39, k32 and k35 from their referenced positions and luminosities. 

In Figure \ref{fig: 3C273}, we see that quasi-stationary knots appear during a specifically long quiescent state of the source of ~2 years. This low activity period is marked by the long decrease of the radio flux seen by OVRO starting from 2011 up to 2013, and also corresponds to an absence of radio ejecta. This observation suggests that the radio-jet of 3C~273 presents at least two quasi-stationary knots (k32, k35) in its nominal (quiescent, non-perturbed) state.

In a few instances, the quasi-stationary knots k32 and k35 seem absent of the observations where one would expect to see them (k35 in two measurements between 2010 and 2012; k32 in one measurement before 2011). 
While these disappearances could be due to observing conditions or instrumental limitations, such as a jet locally outshined by the moving component; they could also highlight a relaxation time for the jet structure to return to its non-perturbed state after the passage of a moving knot.
Linear extrapolations of the motions of the ejecta show that the jet radio luminosity starts to increase when ejecta are emerging from the radio core. 
For each new ejecta, the OVRO flux increase pursues up to the first standing knot marked by k35 ($0.36 \pm 0.08$ mas to the core), and then it decreases while ejecta continue their propagation. VLBA observations of the flux from individual ejecta show a similar behavior as OVRO.
The observed link between VLBI jet kinematics and the radio flux variability in 3C~273 leads to a consistent picture that is in agreement with the scenario we are proposing with our simulations. \\
Firstly, there is a systematic association between the passage of ejecta through the first standing knot k35 and a large flux increase of the overall jet radio luminosity. This crossing is the main phenomenon triggering the radio variability of the source. In addition, the flux increase of the ejecta up to k35 matches what is expected if k35 is a marker of a strong recollimation shock, as the ejecta should undergo a strong acceleration that enhances its Doppler factor by its passage through a large rarefaction zone before this recollimation shock. \\
Secondly, the ejecta enter in an uninterrupted cooling phase after their passage through k35. This is expected when considering that a rarefaction zone should follow the standing shock k35. The second stationary knot marked by k32 does not appear to play an important role for the variability. This suggests that this second recollimation shock is much less powerful than the first one.\\
Finally, the absence of flare can be linked with the passage of the ejecta through the radio core. This may suggest that the radio core marks the jet expanding funnel rather than a first stationary shock (Fig. 4\textit{a} in \cite{Daly_1988}). However studying the variability at higher energy, outside the synchrotron-self-absorbed frequencies, would be necessary to confirm this assumption.

\begin{figure}[h!]
\centering

    \includegraphics[width=\columnwidth]{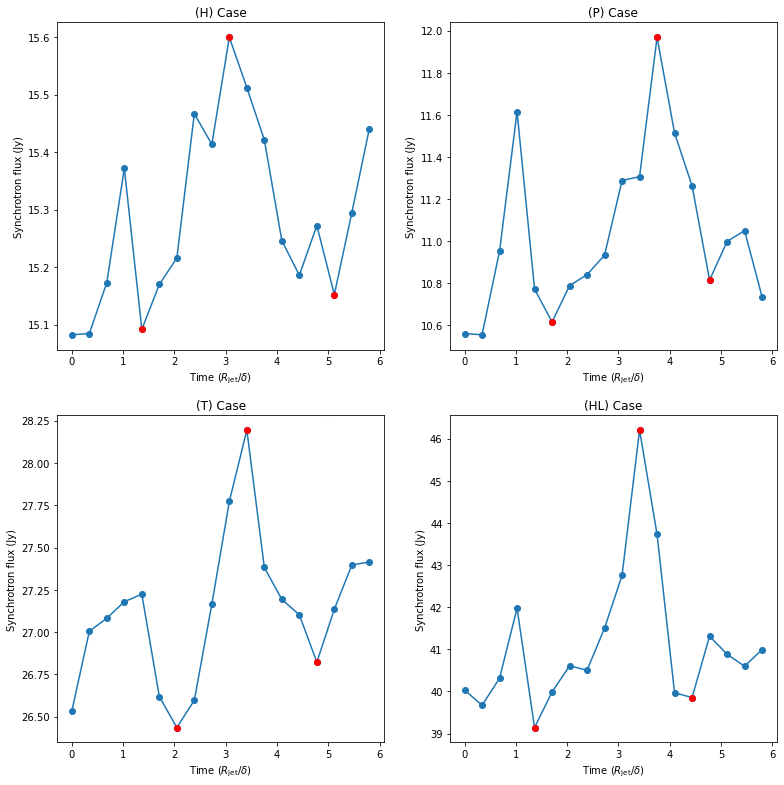}
    
    \caption{Light curve obtain by integrating the total synchrotron flux emitted during the first three interactions between the moving shock and the standing shocks. The computation of the light curve is realized for the four different cases of jets (\textbf{H}, \textbf{T}, \textbf{P} and \textbf{HL}) for $\delta(\theta_{\rm obs} = 2 \degree) \simeq 18$ and $\nu = 15.3 ~\rm{GHz}$. The red dots represent the estimated beginning, maximum and end of the flare.}
    \label{fig: CL results 88_10}
\end{figure}

\begin{figure}
    \centering
    \includegraphics[width =\columnwidth]{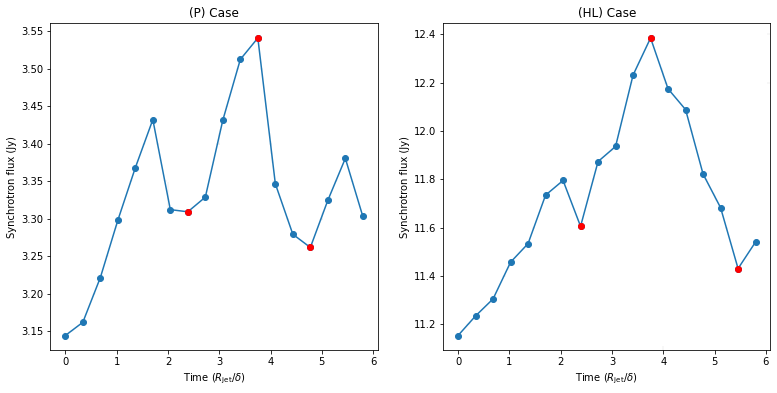}
    \caption{Light curve obtained as in Fig.~\ref{fig: CL results 88_10}. The computation of the light curve is realized for the cases where significant flares were present (\textbf{P} and \textbf{HL}) for $\delta(\theta_{\rm obs} = 6 \degree) \simeq 10$ and $\nu = 15.3 ~\rm{GHz}$. The red dots represent the estimated beginning, maximum and end of the flare.}
    \label{fig: CL results 84_10}
\end{figure}

\section{Discussion}
\label{sec : discussion}

\subsection{Effects of jet stratification and magnetic field structure}

As seen in previous studies, in hydrodynamic \citep{Gomez_1997, Fromm_2018} and magnetized jets \citep{Mizuno_2015, Porth_2015, Fuentes_2018}, we find the well known diamond structure of standing shocks with a clear succession of shock and rarefaction zones in an over-pressured jet. 
A regular standing-shock structure with a linear intensity decrease over the distance from the base is observed. We found a combined effect of the large-scale magnetic configuration and jet stratification on the details of the shock structure. As observed in \cite{Hervet_2017}, the jet stratification induces a larger variety of standing-shock structures due to the interferences between the characteristics waves from the two layers of the jet. 
The toroidal magnetic field strengthens the standing shocks by inducing intense rarefaction regions, where the plasma is strongly accelerated (Fig. \ref{fig: 2D contour energy, Lorentz factor - H T P HL} (top right)). 
Stratification leads to the development of Rayleigh-Taylor instabilities at the outer-inner jet interface and along the standing shock wave as observed in \cite{Toma_2017} in the hydrodynamic case.
In the poloidal and helical cases, the magnetic pressure along the $Z$-axis amplifies these instabilities along the jet. They grow and heat both jet components. Within the inner jet, these instabilities interfere with the standing shocks and lead to a smoother structure and the appearance of a turbulent region at large distance (see Fig. \ref{fig: 2D contour energy, Lorentz factor - H T P HL ejecta} (bottom)).

In a transverse structured jet, the presence of a structured magnetic field amplifies the jet opening angle compared to the hydrodynamic case (Fig. \ref{fig: theta tri}). This is especially apparent in the poloidal and in the helical cases. 
Even if an outer jet layer shields the inner part from a homogeneous ambient medium \citep{Porth_2015}, the magnetic field modifies the topology of the characteristic waves of the fluid.
The presence of a toroidal magnetic field component tends to limit this transverse expansion as observed in \cite{Mizuno_2015} and at large distances in our simulations. On the other hand, the poloidal magnetic field component induces instabilities as we saw; these instabilities lead to a jet decollimation at medium and large distances. 
It was shown by \cite{Marti_2015} that introducing an azimuthal velocity component of the jet flow leads to a centrifugal force that further increases the jet opening. This effect is not currently treated in our models, but we expect that it may have an impact on the standing shock structure, which we will evaluate in a future study. 

\subsection{Ejecta and associated flares in the light curve}
\label{subsect : Ejecta and associated flares in the light curve}

In all cases, the moving shock interacts with the successive standing rarefaction zones, where it is accelerated adiabatically, and collides with standing shocks, where it is heated and decelerated. These interactions lead to the appearance of flares in the light curves, due to thermal energy increase \citep[cf.\ also ][]{Gomez_1997, Fromm_2016}.
The presence of a toroidal component of the magnetic field ensures the cohesion of the moving shock and of the standing shocks. The fact that the moving and standing shock zones are very compact is reflected in the emission of intense and clearly marked flares. We recover a similar behavior in the hydrodynamic case.
On the contrary, where a poloidal field is present, the interaction between the moving shock wave and the more diffuse steady shock zones is weaker and its associated flares are less pronounced.

As we saw, the large variety of flares obtained is related to the intensity of the different knots, stemming from the combination of the characteristic waves of the plasma. 
The outer jet component allows interferences between the stationary shock waves in the inner jet and those of the outer jet. For certain conditions, the two stationary shock  waves can combine and lead to a particularly strong emission.
In the toroidal case, this effect leads to a strong standing shock, with an important rarefaction zone behind it (cf. the fourth standing shock in our simulations). This standing shock region is linked to the luminous flare emission in the light curve (Fig. \ref{fig: CL results 90}). After this interaction, the ejecta is strongly accelerated in the rarefaction zone and will lose its cohesion due to fast adiabatic expansion. 

\subsection{Temporal flare structure}
\label{subsect : Temporal flare structure}

The simulated flares in \textbf{H}, \textbf{T} and \textbf{P} cases are characterized by a temporal asymmetry with a fast rise and a slower decay phase, even though this is not always clearly visible due to the varying strength of the standing shocks and the limits in temporal resolution. 
When the moving shock interacts with the standing shocks, it heats and compresses them. Afterwards, the knots decompress. These interactions induce the formation of trailing recollimation shocks, already observed by \cite{Agudo_2001} and \cite{Mimica_2009}, which will perturb standing knots along the jet axis and make them oscillate.
This is associated with remnant emission from the shocked knots during their adiabatic cooling phase. This process is observed in all  cases, but most clearly in the cases with pronounced interaction between the moving shock and the standing shock (cases {\bf H} and {\bf T}). Indeed, after the strong flare, we see in the toroidal case a delay between the emission of the ejecta and the emission of the jet (Fig. \ref{fig: CL results 90}). This is the emission signature of the shocked knot, which causes a slight asymmetry. 
This additional radiation counterpart will tend to soften the slope after the ejecta has passed. This behavior is in accordance with the observed flare structure described by \cite{Hovatta_2008, Nieppola_2009} and obtained numerically from over-pressured jets by \cite{Gomez_1997, Fromm_2016}. 

\subsection{Light curve with small observation angle}
\label{subsect : Light curve with small observation angle}
 As the angle decreases to $15 \degree$ or $2 \degree$ (Fig. \ref{fig: CL results HL}), the effect of the absorption of the moving shock along the line of sight becomes more important at low frequencies. Thus, the observed flare intensity, duration and asymmetry decrease as the occultation by the moving shock becomes more important. 
 Moreover, after the flare associated with the interaction of the moving shock with the most intense steady knot, the flux intensity decreases. This is also observed by \cite{Fromm_2016} but, in our case, we have not taken into account the effect of the "light travel delay" of photons emitted in different parts of the jet, which would lead to smoother light curves with a longer decay. This behavior is well distinguishable with viewing angle $\theta_{\rm obs}=2 \degree$. 

\subsection{Comparison with observation of 3C 273}
\label{subsect : Link with 3C 273}

With our shock propagation model, we arrive at a consistent, for now qualitative interpretation of the flaring behavior observed in 3C~273.
Detailed modeling of the observed light curve is out of the scope of this work and will be addressed in a future publication. 
The number of recollimation shocks present in our simulated jets is much higher than the number of standing knots observed in the case of 3C~273, mostly
due to the purely cylindrical shape of our jets. In real jets that are imperfectly aligned, the superposition of knots along the line of sight and radio opacity might 
also lead to an obscuration of standing knots for the observer.
VLBI observations of the most nearby radio-galaxy (M\,87, \cite[e.g.,][]{Asada_2014}) and of other AGN \citep{Hervet_2016} show the presence of multiple standing features that may be interpreted as series of recollimation shocks.
Before the moving shock waves interact with the first standing shock, extrapolation of the emitted flux seems to indicate that they are undergoing a strong acceleration. This phenomenon appears clearly in our results, with the moving shock passing through the first rarefaction zone before interacting with the first knot. The acceleration seems especially pronounced in the magnetized cases as shown in the Fig. \ref{fig: lfac max ej}. 
After the passage of moving shock waves, the disruption of the standing knots can be explained by the dynamics and / or by the radiative transfer within the jet. 
Concerning the dynamics aspect, the trailing recollimation shocks can perturb momentarily the standing-shock structure, as we saw in our simulations. On the other hand, the apparent "disappearance" of standing knots, if confirmed, may suggest that they are simply obscured by the brighter moving knots (k31 or k37) that hide the quasi-stationary knot, like we see in our synthetic light curve. 

In our model, the light curve (Fig. \ref{fig: CL results 88_10}) obtained at frequency 15.3 GHz (OVRO/MOJAVE frequency) and viewing angle $2\degree$ \citep[e.g.,][]{Hervet_2016} shows the interaction of the moving shock with the first two stationary shocks. The light curve integrates the flux emitted by the whole jet like a single dish telescope. We can thus compare our results with the OVRO data, in particular with the k37 event, which is isolated from other events. \\
We should note that the observations from \cite{Jorstad_2017} found a different viewing angle of $6.4 \pm 2.4 \degree$. To investigate the effects of the viewing angle, which is not precisely known, we also compute the light curves for $6 \degree$ (Fig.\ref{fig: CL results 84_10}). \\

In the simulations, as in the observations, we find a flare during each interaction between the moving shock wave and a standing shock. However, the typical flare duration is very different ($\sim 800$ days for k37 and $\sim 150$ days for our flares on average). This could be due to differences in the morphology of the jet, the size of the ejecta and stationary knots, and the uncertain value of the observation angle of 3C~273.  \\
The shape of the observed flares seems to show a small asymmetry. To quantify this effect, we used the method proposed by \cite{Roy_2018, Nalewajko_2013} to compare the doubling (or halving) time in the rise (or fall) phase of the flare. Applying the method to the k37 flare, we found $\xi_{\rm k37} = 0.12 \pm 0.03$ where the fall time is superior than the rise time. Applying the same method to the simulated flares, we found respectively $\xi_{\rm H} = 0.14 \pm 0.01$, $\xi_{\rm T} = 0.13 \pm 0.02$, $\xi_{\rm P} = 0.12 \pm 0.03$  and $\xi_{\rm HL} = -0.29 \pm 0.06$ for the  hydrodynamic, toroidal, poloidal and helical case for $\theta_{\rm obs} = 2\degree$ (Fig. \ref{fig: CL results 88_10}). \\

At $\theta_{\rm obs} = 6\degree$, the shape of the light curve changes significantly due to beaming effects, such that the partial superposition of the first three flares does not allow us to clearly determine the presence or absence of asymmetries. The peak of the second flare is only well visible in the \textbf{P} and \textbf{HL} cases (Fig. \ref{fig: CL results 84_10}). In the \textbf{HL} case, no asymmetry is found, while in the \textbf{P} case it has switched signs compared to the simulation at $\theta_{\rm obs} = 2\degree$ (we found for this case $\xi_{\rm P} = -0.23 \pm 0.03$). We should also note that the current version of our model does not take into account time delay effects, which may play an important role at small angles. A more detailed study is beyond the scope of this work and will be applied in a future study to dedicated simulations for a given data set.\\

The amplitude of the flares is on average much larger in 3C~273 where the variability in radio can reach around twice the baseline value, compared to an increase of $\sim 15\%$  in the same-frequency, small angle simulations (Fig. \ref{fig: CL results 88_10}). This difference depends on the same characteristics as the flare duration. 
\\
For the general picture, the main difference with 3C~273 is the presence of only two visible recollimation shocks, where only the first one is linked to strong flares, compared to a greater number of standing shocks and resulting flares in our simulations.
The number of knots observed is however strongly linked to the angular resolution and sensitivity of the VLBA.
Another stationary knot very close to the core at $\sim 0.10-0.16$ mas, noted A1, ST1, or S1 has been detected when observing the jet at 43.2 GHz \citep{Jorstad_2005, Savolainen_2006, Lisakov_2017,Kim_2020}.

As discussed in Section \ref{subsect : Ejecta and associated flares in the light curve}
for the simulations, this effect may be explained in 3C~273 by a strong interaction between shock waves in the inner jet and outer jet occurring at the position of k35, at 0.36 mas to the core. This zone can lead to a major outburst when interacting with a moving shock. It also has the specificity of disrupting the downstream shock structure, which would explain the weak presence of the downstream stationary shock k32, as well as the absence of significant flare event associated with it, and the apparent adiabatic cooling of ejecta moving outside of k35. 

To directly model observed flares with our radiative SRMHD code, a more important opening of the jet, reflecting changes in the density profile of the ambient medium, will be required. This should lead to a shock structure dominated by a few standing shocks close to the core. An implementation of the "light travel delay" effect \citep[cf.]{Chiaberge_1999} and of radiative cooling will also be needed for a more realistic description of the radiative emission.

\section{Conclusions}
\label{sec: conclusion}

We have investigated the effect of the large-scale magnetic field on the standing shocks and their interaction with ejecta within a two-component relativistic jet. The associated light curves, which were computed at two radio frequencies ($\nu = \left( 1,~15.3\right)~ {\rm GHz}$) and for several observation angles ($\theta_{\rm obs}=\left(2\degree,~15\degree,~90\degree\right)$), show a variety of flares with varying durations and amplitudes.

Two-component magnetized jets are characterized by a complex standing-shock structure due to the interaction of characteristic waves propagating in the two jet components. In this way, jet stratification leads to the appearance of radio knots with a range of intensities along the jet. This is especially apparent in the toroidal case, where we recover a strong standing shock, giving rise to a pronounced flare in the interaction with the moving shock wave. 
Temporal asymmetry associated to the relaxation phase of the shocked standing knot is well visible for the strongest flares.
The introduction of large-scale magnetic fields is seen to cause an intrinsic opening of the jet with an opening angle up to three times larger than the hydrodynamic case for our jet configuration. 

Our scenario of moving ejecta interacting with standing shocks inside a two-component jet provides a good description of the kinematics and light curves seen in the jet of the FSRQ type blazar 3C~273 with VLBI and single-dish radio observations. In a preliminary study, at observation angle of $2\degree$, an asymmetry in the simulated flare profiles, with a fall-time that is longer than the rise-time, was seen for the hydrodynamic, toroidal and poloidal cases, consistent with what is observed in the OVRO data for this source.

\begin{acknowledgements}
The authors thank the anonymous reviewer for his peer review. The computations of the SRMHD results were carried out on the OCCIGEN cluster at CINES\footnote{\url{https://www.cines.fr/}} in Montpellier (project named \texttt{lut6216}, allocation \texttt{A0050406842}) and from the MesoPSL Cluster at PSL University\footnote{\url{http://www.mesopsl.fr/}} in the Observatory of Paris. This research has made use of data from the MOJAVE database that is maintained by the MOJAVE team \citep{Lister_2018}, and from the OVRO 40m monitoring program which was supported in part by NASA grants \texttt{NNX08AW31G}, \texttt{NNX11A043G} and \texttt{NNX14AQ89G}, and NSF grants \texttt{AST-0808050} and \texttt{AST-1109911}, and private funding from Caltech and the MPIfR. The authors wish to thank M. Lister for providing preliminary MOJAVE data points for this study. We wish also thank M. Lemoine for providing insightful discussion during the study.   
\end{acknowledgements}

\bibliographystyle{aa.bst} 
\bibliography{2021_01_flare.bib} 

\onecolumn

\begin{appendix}

\section{Additional Figures}

\begin{figure}[h]
    \centering
    \subfloat{{\includegraphics[width=0.48\columnwidth]{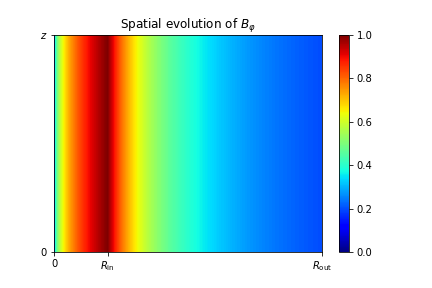}}}%
    \qquad
    \subfloat{{\includegraphics[width=0.48\columnwidth]{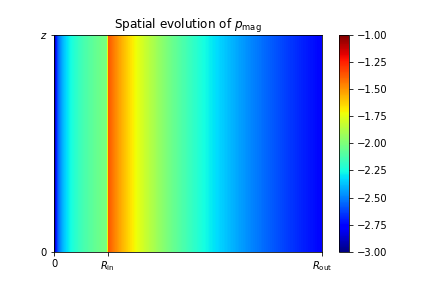}}}%
    \caption{\textit{Left}: Initial representation of the toroidal magnetic field strength in space following Eq. \ref{Eq:SubS_Model_Bphi} and assuming $\left(B_{\rm \varphi,in,0},\,B_{\rm \varphi,out,0}\right) \equiv 1$. \textit{Right}: Initial representation of the magnetic pressure following Eq. \ref{eq: thermal pressure}.}
    \label{fig: Additional figs}
\end{figure}

\end{appendix}
\end{document}